\documentclass{article}
\usepackage{fullpage}
\usepackage{tikz}
\usepackage{hyperref}
\usepackage{amsmath}
\usepackage{amsthm}
\usepackage{booktabs}
\usepackage{graphicx}
\usepackage{subcaption}
\usepackage[normalem]{ulem}
\theoremstyle{definition}
\usepackage{multirow}
\usepackage{array} 
\newcolumntype{P}[1]{>{\raggedright\arraybackslash}p{#1}}

\newcommand\shortsection[1]{\vspace{6pt}{\noindent\bf #1.}}

\newcommand{\originalviews}[1]{\texttt{original\_views}}
\newcommand{\dpviews}[1]{\texttt{dp\_views}}
\newcommand{\roundedviews}[1]{\texttt{rounded\_views}}

\usepackage[backend=biber,sorting=nyt,sortcites,maxbibnames=99]{biblatex}
\usepackage{biblatex}
\begin{document}

\title{``Having Confidence in My Confidence Intervals'': How Data Users Engage with Privacy-Protected Wikipedia Data\thanks{H.T. and J.S. contributed equally to this work as first-authors. R.C., G.K., S.K., and E.M.R. contributed equally to this work as last-authors.}}

\author{Harold Triedman\thanks{Cornell Tech. Email: \texttt{hjt36@cornell.edu}. H.T. supported by an NSF Graduate Research Fellowship.} \and 
Jayshree Sarathy\thanks{Northeastern University. Email: \texttt{j.sarathy@northeastern.edu}. J.S. supported in part by NSF grant ITE-2429841.}
\and 
Priyanka Nanayakkara\thanks{Harvard University. Email: \texttt{priyankan@g.harvard.edu}. P.N. supported in part by NSF grant ITE-2429838.}
\and
Rachel Cummings\thanks{Columbia University. Email: \texttt{rac2239@columbia.edu}. R.C. supported in part by NSF grants
CNS-2138834 (CAREER), EEC-2133516, and ITE-2429841, and by DARPA (contract number W911NF-21-1-0371). Any opinions, findings, and conclusions or recommendations expressed in this material are those of the authors and do not necessarily reflect the views of the United States Government or DARPA.}
\and
Gabriel Kaptchuk\thanks{University of Maryland. Email: \texttt{kaptchuk@umd.edu}. G.K. supported in part by NSF grant ITE-2429839.}
\and
Sean Kross\thanks{Fred Hutch Cancer Center. Email: \texttt{skross@fredhutch.org}. S.K. supported in part by NSF grant ITE-2429840.}
\and Elissa M. Redmiles\thanks{Georgetown University. Email: \texttt{er913@georgetown.edu} E.M.R. supported in part by NSF grant ITE-2429838.
}}

\date{}

\maketitle

\begin{abstract}
In response to calls for open data and growing privacy threats, organizations are increasingly adopting privacy-preserving techniques such as differential privacy (DP) that add noise when generating published datasets. These techniques are designed to protect privacy of data subjects while enabling useful analyses, but their reception by data users is underexplored. We developed documentation that presents the noise characteristics of two Wikipedia pageview datasets: one using rounding (heuristic privacy) and another using DP (formal privacy). After incorporating expert feedback ($n=5$), we used these documents to conduct a task-based contextual inquiry ($n=15$) exploring how data users—largely unfamiliar with these methods—perceive, interact with, and interpret privacy-preserving noise during data analysis.

Participants readily used simple uncertainty metrics from the documentation, but struggled when asked to compute confidence intervals across multiple noisy estimates. They were better able to devise simulation-based approaches for computing uncertainty with DP data compared to rounded data. Surprisingly, several participants incorrectly believed DP's stronger utility implied weaker privacy protections. Based on our findings, we offer design recommendations for documentation and tools to better support data users working with privacy-noised data.
\end{abstract}

\maketitle

\section{Introduction}

Data has become a critical and valuable resource over the last two decades. Organizations have been encouraged to open up access to their data for external researchers and the public. At the same time, however, privacy researchers have demonstrated potential for attacks on open data and aggregate statistics \cite{sweeneySimpleDemographicsOftena}. Theoretical results (e.g. \cite{dinurRevealingInformationPreserving2003}) and real-world attacks (e.g. \cite{narayanan2008robust, USENIX:Cohen22, abowdSimulatedReconstructionReidentification2025}) have repeatedly demonstrated that the release of too many, too accurate statistics enables analysts to reconstruct sensitive datasets and identify individuals within them. Such attacks can not only cause real harms to data subjects, but also threaten the reputation of organizations that release data.

To mitigate these privacy threats while maintaining the spirit of open access, organizations have started to incorporate \emph{privacy noise} when generating datasets for publication \cite{khavkin2025differential, desfontainesListRealworldUses,nanayakkara2025practitioners}. Some noising techniques, such as rounding and swapping~\cite{dalenius1982data}, have been used for decades; others like differential privacy (DP)~\cite{dwork2006calibrating} are more recent technical developments. The goal of these noising techniques is to protect the privacy of individual data subjects while preserving overall patterns in the data.

Real-world data releases that include privacy noise have highlighted the struggles and anxieties of data users within this new paradigm. The most prominent example followed the US Census Bureau’s announcement that they would use DP to release 2020 decennial census\footnote{We will use lowercase 'census' to describe data products and uppercase 'Census' to describe the US Census Bureau.} data~\cite{hansen2018reduce}, which was met with shock from data users. Despite various efforts to communicate with data users and provide resources explaining DP~\cite{SolicitingFeedbackUsers2018, bureauBalancingActProducing}, tensions continued to grow within data user communities who felt that the addition of noise for privacy reasons was unnecessary and would compromise data quality~\cite{hotz2022chronicle}. These heated debates continued long after the data were released, resulting in lawsuits~\cite{alabama_lawsuit,scheinderCensusAP2025}, news articles (e.g. \cite{wezerekChangesCensusCould2020,winesCensusNYT2020}) and studies (e.g. ~\cite{boyd_differential_2022,cohen2022private,steed2022policy,christ2022differential}) calling for more attention to how data users' analyses would be impacted by changes to the data. Other high-profile data releases, such as Facebook's Social Science One initiative that provided social media viewing data to researchers~\cite{socialscienceoneOurFacebookPartnership}, generated similar, albeit less public, dissatisfaction from data users.

The rocky reception to these data releases demonstrates the privacy community's gaps in understanding data users' perceptions of noisy data and best practices for communication.
Prior work has focused on
communication of privacy guarantees to data subjects (e.g., \cite{fischer2024overview, bullek_towards_2017, xiong_towards_2020, cummings_i_2021, franzen_am_2022, karegar_exploring_2022, smart2022understanding, nanayakkara_what_2023, smart2024models}), guidance for data curators who wish to apply privacy protections when releasing sensitive data (e.g. \cite{gaboardi2016psi,murtagh2018usable,dwork_differential_2019,munillagarridoLessonsLearnedSurveying2023}), and usability interventions for software developers and analysts who aim to implement privacy-preserving mechanisms (e.g. \cite{nanayakkara_visualizing_2022,panavas2023investigating,panavas2025illuminating,ngong2024evaluating,song2024inherently, nanayakkara2024measure}). However, the experiences of data users who use \emph{already}-privacy-noised datasets are less studied.
Failing to address this perception and communication gap puts the broader project of opening access to data, while protecting data subjects' privacy, at risk.

In this work, we explore the perspectives of data users\footnote{For this study, we consider data users who are experienced with quantitative data analysis and have familiarity with analyzing social or web datasets.} contending with noise in data releases using datasets from the Wikimedia Foundation (WMF), which has released sensitive data using two different methods of privacy noise: rounding and DP. WMF is the only organization we are aware of\footnote{The US Census Bureau used two different methods of privacy protection, swapping and differential privacy, for its 2010 decennial census and 2020 decennial census, respectively. These two datasets are similar but not identical. } that has released the same dataset using different noising methods, which provides a natural opportunity to understand how data users engage with heuristic versus formal privacy noise techniques in datasets. In addition, it enables us to study data users' perceptions in less political and adversarial dynamics than with the 2020 census and Facebook Social Science One releases.

\begin{quote}
\textbf{RQ:} How do data users perceive, interact with, and interpret noise injected into data for privacy protection?  Does the nature of the noise (rounding vs. DP) impact data users' engagement with the data? 
\end{quote}
To explore these questions, we conducted a task-based analysis study with 15 participants experienced in data science. Each study session included a contextual inquiry using data analysis tasks, as well as a semi-structured interview. 
To ground our interviews in real-world analysis environments, we provide participants with dataset documentation, which is critical for enabling data users to responsibly work with data~\cite{mitchellModelCardsModel2019a,gebru_datasheets_2021,kelley2009nutrition} and aligns with typical documentation practices for WMF data releases~\cite{WMF_analytics_api}. 
We design documentation for each of the datasets, incorporating feedback from five experts in privacy communication, and provide these documents to participants during the study. 

We find that participants were well-versed in existing sources of error in non-privacy-protected Wikipedia datasets. They 
were more easily able to understand the process of rounding, but they felt better equipped to devise simulations and develop empirical confidence intervals using the DP data. 
For both rounding and DP, however, participants struggled to understand how uncertainty would scale across multiple perturbed data points.
Participants had mixed and surprising perceptions about the privacy protections offered by rounding and DP.
We conclude with recommendations to help data users work effectively with privacy-noised datasets, such as building tools to automatically compute confidence intervals and track uncertainty in downstream analyses. We point to the need for future work to better understand how different audiences perceive the relationship between privacy and accuracy of noisy data.

\section{PRELIMINARIES AND RELATED WORK}
\label{sec:related-work}

\subsection{Technical background on rounding and DP}
\label{sec:background}

In this work, we focus on two approaches towards privacy-preserving noise addition: the first is a common, general approach of \emph{rounding}, and the second is a more recently developed, mathematically formal approach of \emph{differential privacy (DP)}. Below, we provide brief, non-technical descriptions of both approaches and discuss our motivations for studying these approaches.

\textit{Rounding} is a statistical disclosure limitation (SDL) technique in which a quantitative variable is replaced by a smaller \textit{rounding set}~\cite{willenborgElementsStatisticalDisclosure2001,domingo-ferrerSurveyInferenceControl2008}. The rounding set typically contains multiples of some base value (for example, 10 or 100). 
Rounding is a heuristic technique for privacy-preserving data releases, meaning that it does not offer provable guarantees of how well individuals in the dataset are protected thereafter. 
Rounding is also commonly used for purposes other than privacy, such as grouping numbers into a smaller set, conveying imprecision, and simplifying the presentation of statistics.

\textit{Differential privacy (DP)} is a formal, mathematical framework for privacy-preserving data analysis that limits the amount of information a statistical release reveals about individuals in the data~\cite{dwork2006calibrating}.
Satisfying DP requires injecting a carefully calibrated amount of statistical noise into the release to protect the privacy of individuals in the dataset. 

We consider rounding and DP as these are the two techniques used by WMF to release versions of the same Wikipedia pageview dataset, as we describe in Section~{\ref{subsec:datasets}}.
DP was initially adopted as an experimental data sharing technique in 2023, and both datasets have continued to be published while WMF decides if and when to deprecate one or the other.

We are additionally motivated to study rounding and DP due to their key differences. Rounding is easy to understand, commonly used, and not specific to privacy protection.
DP requires more explanation and technical background, is less commonly used, and pertains to a specific, mathematical definition of individual privacy.
Regardless of the technique used, there is fundamental tradeoff between privacy and utility: the release of too many, too accurate statistics degrades the privacy afforded to individuals in the dataset, and injecting noise into the data necessarily introduces some loss in accuracy~\cite{dinurRevealingInformationPreserving2003}.

We describe rounding and DP as privacy-preserving, noise addition mechanisms in order to align with how WMF initially conceived of these methods in the implemented data releases.\footnote{Some experts, including one we interviewed for this study, reject the notion that rounding should be considered a method for privacy protection given its heuristic nature. Others may not consider rounding to be a noise addition mechanism, but rather a tool for introducing or conveying imprecision. For this paper, we align the framing of rounding with how it was considered by WMF in their real data deployment, and we consider introducing imprecision to be synonymous with adding noise. See the following WMF community threads which discuss applying privacy protections to the data: \url{https://phabricator.wikimedia.org/T207171\#6591900}, \url{https://phabricator.wikimedia.org/T207171\#6592952}, \url{https://phabricator.wikimedia.org/T207171\#6614729}} 
This framing also differentiates rounding and DP from other SDL techniques that limit or suppress the amount of data that is published.
Aligning our study with the real WMF data releases provides a natural opportunity to understand data users' engagement with heuristic versus formal privacy noise in datasets. 

The technical aspects of injecting privacy noise into data have been well-studied over the last several decades~\cite{dwork2014algorithmic,vadhan2017complexity,cowan2024hands}. More recently, scholars have explored the real-world impacts and human aspects of privacy-preserving data analysis. We discuss some of this work in the following sections.

\subsection{Releases of privacy-preserving data}
Prior data releases have demonstrated a pressing need to understand data users’ engagement with privacy-noised datasets. One key example is the US Census Bureau's release of DP data for the 2020 decennial census. Responding to scientific consensus around new privacy threats and an internal demonstration attack on 2010 census data that highlighted the weakness of heuristic privacy protections of swapping~\cite{abowdSimulatedReconstructionReidentification2025}, the Census Bureau announced in 2018 that they would release the 2020 decennial data using DP~\cite{hansen2018reduce}. What seemed like a natural modernization of privacy protections for census data resulted in years of negative attention and controversy for the Census Bureau, including strained relationships with core data user communities~\cite{hotz2022chronicle}, critical news headlines \cite{wezerekChangesCensusCould2020,winesCensusNYT2020}, and lawsuits~\cite{alabama_lawsuit,scheinderCensusAP2025}. 

Several scholarly works have analyzed what went wrong in the 2020 census release.
One key insight is that the transparent addition of noise using DP upended data users’ mental models around census data as fact~\cite{boyd_differential_2022}. Researchers highlighted how core uses of census data in redistricting and funding formulas do not fully account for uncertainty in census data, and thus are not robust to error from both privacy noise as well as from the data collection and processing pipeline~\cite{cohen2022private,steed2022policy}. Although noise addition via DP had a smaller accuracy impact than the prior heuristic forms of privacy protection (i.e., swapping~\cite{christ2022differential}), data users still felt that the use of DP was an existential threat to the utility of these critical data products~\cite{ruggles_differential_2019}. Communication efforts between the Census Bureau and data users failed to bridge these gaps~\cite{hawes2020implementing,hotz2022chronicle,abdu2024algorithmic,nanayakkara2022s}.

These works emphasize that the field does not have a good understanding of how data users perceive and interact with data that contains privacy protections, and how data users' engagement differs based on contextual factors including the types of privacy protection that were used~\cite{oberski2020differential}. Some works have advanced theoretical arguments about how modernizations to census data collection impact the public’s reception~\cite{sarathy2025statistical}, but these do not empirically explore the nature of data users’ engagement with different privacy protections.

Other deployments of DP, such as via Facebook’s Social Science One project \cite{socialscienceoneOurFacebookPartnership} which provided DP statistics about social media activity to researchers, faced different but similarly negative reactions from data users. These data users were experts in quantitative social science research and adept at incorporating uncertainty into their analyses, but they were unable to verify that the data was complete (which, it turned out not to be \cite{silvermanFundersAreReady2019}), and struggled to analyze the impact of privacy-preserving noise addition on their downstream analyses and data transformations~\cite{evans2023statistically}. This example indicates that the different contexts of data release, and different technical expertise of data users, creates diverse challenges for designing usable data releases.

\subsection{Studying human interactions with privacy-preserving data}
\label{subsec:related-work:user-interactions}
There has been a growing focus on human aspects of privacy-preserving data analysis. Much of the work in this area focuses on communication of privacy guarantees to data subjects (e.g., \cite{fischer2024overview, bullek_towards_2017, xiong_towards_2020, cummings_i_2021, franzen_am_2022, karegar_exploring_2022, smart2022understanding, nanayakkara_what_2023, smart2024models}), tools for data curators who wish to release privatized data (e.g., \cite{gaboardi2016psi,murtagh2018usable, nanayakkara_visualizing_2022,sarathy_dont_2023}), or usability interventions for software developers and analysts who are implementing privacy-preserving mechanisms (e.g., \cite{panavas2023investigating,panavas2025illuminating, nanayakkara2024measure,song2024inherently}). Less studied are the experiences of data users who use datasets that have already been privacy-protected. We briefly summarize these three lines of work below. 

The first line of work focuses on challenges of usable privacy for data subjects. 
There is a long HCI literature in this space~\cite{fischer2024overview}, but most relevant to our work are those that study methods for explaining statistical privacy guarantees to data subjects. These works explore the effectiveness of various textual and visual explanations to characterize differential privacy threat models and privacy guarantees~\cite{bullek_towards_2017, xiong_towards_2020, cummings_i_2021, franzen_am_2022, karegar_exploring_2022, smart2022understanding, nanayakkara_what_2023, smart2024models,dibiaWeNeedStandard2025}. We take inspiration from these prior works but consider the perspectives of data users rather than data subjects, and focus on explanations of data utility rather than privacy protections.

The second stream of work focuses on usability challenges for data curators, who are organizations and researchers that collect data from data subjects, or repositories that ingest datasets and manage access and releases (such as Dataverse~\cite{magazine2011dataverse}). Early work focused on building interactive software tools and systems~\cite{gaboardi2016psi,murtagh2018usable,holohan2019diffprivlib,berghel2022tumult} to help data curators add DP noise to datasets before publishing them. Others have explored the challenges that privacy developers, data curators, and other stakeholders face when deploying DP~\cite{dwork_differential_2019,agrawal2021exploring,munillagarridoLessonsLearnedSurveying2023,rosenblatt2024data,nanayakkara2025practitioners} and have evaluated the usability of open-source tools and libraries for DP~\cite{ngong2024evaluating,song2024inherently}.
Researchers have also highlighted the need for visualizations to guide data curators in this process and designed systems that illustrate the impact of parameter selection—such as setting the \emph{privacy-loss parameter}, $\varepsilon$, for DP—on the privacy-utility tradeoff~\cite{nanayakkara_visualizing_2022,panavas2023investigating,panavas2025illuminating}.

Finally, recent releases of DP data have highlighted the need to consider not just the perspectives of data subjects and curators, but also data users who receive privacy-protected datasets and must contend with privacy noise in their analyses~\cite{oberski2020differential,boyd_differential_2022,williams2024disclosing}.
This perspective has been understudied in the literature, with limited prior works that interrogate data users' interactions and perceptions of privacy-noised data. 
Williams et al.~\cite{williams2024disclosing} conduct a survey of economists to understand how DP aligns with their expectations for usable data. They find that economists are concerned about the addition of noise and have low tolerance of error introduced from privacy noise, and their survey results motivate our inquiry into the point at which data users interact with noisy data. 
Sarathy et al.~\cite{sarathy_dont_2023} consider usability challenges of DP tools from the perspective of data users, but their study focuses on an \emph{interactive} context where analysts query for DP-noised statistics via an interface. They find that data users have a hard time interpreting DP results and error in this dynamic query context, as users lack access to raw data to ground their findings and identify potential problems. However, this finding does not speak to data users' experiences in a \emph{non-interactive} setting, where data users are given data that has been privacy-protected already and can query this data freely. The non-interactive setting is much more common in real-world deployments of privacy-preserving analysis techniques, and is therefore the focus of this work.

\subsection{Data and model documentation}
\label{subsec:model-documentation}

HCI and critical data studies have emphasized the need for robust documentation of data and models. Prior work in these areas have provided guidelines for documenting datasets~\cite{gebru_datasheets_2021, benderDataStatementsNatural2018,holland2020dataset} and AI and ML models~\cite{mitchellModelCardsModel2019a, crisanInteractiveModelCards2022a}. These works advocate for standardized procedures to communicate information such as data provenance, intended use cases, model performance, and potential pitfalls around using the data or model. 

In the usable privacy literature, prior work has focused on documentation to convey privacy protections to data subjects. Privacy nutrition labels use food nutrition labels as a metaphor for structuring documentation on data collection, retention, sharing, and processing activities~\cite{kelley2009nutrition}. This documentation is deployed in real-world contexts, such as in the Apple App Store~\cite{li2022understanding}). Recent work \cite{dibiaWeNeedStandard2025} has extended this notion to differentially private data processes. However, these techniques focus on explaining privacy guarantees to data subjects, and do not extend to explaining data utility to data users. 

In short, developing effective documentation for data users poses new benefits and challenges for communication that are not addressed in prior work~\cite{cummings2023centering}. For example, explaining privacy guarantees in a simple way may be less of a priority, as data users do not make decisions based on the privacy guarantees in the way data subjects do. At the same time, there is an additional need to convey the impact of privacy noise on utility metrics (which may be highly context-dependent) in a way that aligns with data users' mental models of data uncertainty and does not overwhelm them with too much technical or statistical information.
In this work, we explore these challenges and provide initial insights towards designing effective documentation about privacy-noised datasets geared towards data users.

\section{Data, Documentation, and Tasks}

Here we describe the datasets, documentation and tasks that grounded our study. These artifacts served as \emph{design probes}~\cite{hutchinson_technology_2003} for understanding data users’ engagement with privacy-noised data, meant to elicit reactions from participants and to serve as a starting point for future iteration and study.

\subsection{Datasets} 
\label{subsec:datasets}
WMF has publicly committed to open data access~\cite{WMF_open_access_policy}, and sometimes publishes datasets requested by various WMF constituencies (e.g., wiki administrators, community builders, or researchers). Participants in our study interacted with three publicly-available datasets published by WMF. 

\paragraph{Global pageview dataset} The first dataset has been published by WMF for decades\footnote{\url{https://dumps.wikimedia.org/other/pageviews}} and contains counts of views to Wikimedia pages, broken down by project (e.g., Spanish Wikipedia or Japanese Wiktionary) and day (e.g., May 1, 2024).\footnote{Although these datasets contain non-Wikipedia projects, participants exclusively filtered them to look at Wikipedia data. Therefore, we refer to them as ``Wikipedia data''.} This dataset is aggregated globally: pageviews to the same page that originate from different countries are added together. We start with this dataset to allow participants to familiarize themselves with the data domain and understand how to conduct basic operations on the data.

\paragraph{Pageviews by country (rounded)} The second dataset has been published by WMF since 2021\footnote{\url{https://doc.wikimedia.org/generated-data-platform/aqs/analytics-api/reference/page-views.html}} and contains pageview counts, broken down by project, day, and country. These pages are noised by rounding the exact pageview count (known in our documentation as \originalviews~) \textit{up} to the next 100 to derive \roundedviews~. All pages-country tuples that receive $\leq$1000 \originalviews~ in a day are excluded from the dataset. These parameters were established as safe heuristics through public discussion among WMF staff and volunteer developers prior to dataset release.\footnote{See, e.g., \url{https://phabricator.wikimedia.org/T207171}}

\paragraph{Pageviews by country (DP)} The third dataset has been published by WMF since 2023\footnote{\url{https://analytics.wikimedia.org/published/datasets/country_project_page/}} and, similar to the rounded dataset, contains pageview counts broken down by project, day, and country. Input rows are filtered to the first ten unique pageviews per device per day. These pageviews are counted, and the counts are noised by a Gaussian mechanism~\cite{dwork2014algorithmic} (which satisfies \emph{zero-concentrated DP}~\cite{bun_concentrated_2016}, a relaxation of the standard DP definition) to derive \dpviews~. Outputs are filtered to only include \dpviews~ $\geq$90. These parameters were chosen by WMF staff through experimentation, in order to ensure that the resulting output dataset has daily error rates below pre-established thresholds. More details about this dataset are discussed in Adeleye et al.~\cite{adeleye_publishing_2023}. 

We trimmed these datasets to a small range of dates to make them easier to load into a Jupyter notebook. 
For the global dataset, we show pageviews on April 21 and 22, 2024 for pages that receive at least 25 views; for the by-country datasets, we show privacy-protected pageviews from May 1-10, 2024 for pages that meet the criteria for rounding (page-country tuples with $> 1000$ \originalviews~) or DP (page-country tuples with $\geq 90$ \dpviews~) as described above. Other than restricting to a subset of dates, we did not change parameters for thresholding and filtering that WMF uses for each dataset in the actual data releases.

\paragraph{Why apply privacy protections?}
It may not be immediately apparent why privacy protections beyond thresholding (e.g. not reporting pages with $\leq T$ \originalviews, for some threshold $T$) 
were considered and implemented by WMF for the pageviews by country datasets. Below, we sketch two attacks to illustrate how thresholding may be insufficient, especially against adversaries with auxiliary knowledge.

Suppose an adversary {$\mathcal{A}$} would like to pinpoint the country location {$C$} of person {$\mathcal{X}$}, given {$\mathcal{X}$}'s email address. {$\mathcal{A}$} can (1) deploy bots to view a rarely-viewed page {$p$} exactly {$T$} times per day per country, for all countries, and (2) email {$\mathcal{X}$} a link to page {$p$}. When the link is opened, $p$'s view count will exceed $T$ and appear in the dataset for country $C$. Thus, {$\mathcal{A}$} can determine with high likelihood that {$\mathcal{X}$} is located in country $C$. With additional auxiliary information, {$\mathcal{A}$} can further deduce {$\mathcal{X}$}'s specific location.
Next, suppose {$\mathcal{A}$} wants to track whether person {$\mathcal{X}$} is reading some controversial Wikipedia pages {$p_1, \ldots, p_n$}, and assume {$\mathcal{A}$} knows that {$\mathcal{X}$} is a minority language speaker in their country $C$ (e.g. a Malaysian speaker in Luxembourg). {$\mathcal{A}$} can deploy bots to view $p_1, \ldots, p_n$ in Malaysian Wikipedia exactly $T$ times per day from country $C$.  
If one of the pages subsequently appears in the dataset, {$\mathcal{A}$} will be able to deduce with high likelihood that {$\mathcal{X}$} viewed that page on that day.

Whether these simple attacks are relevant or worth protecting against are questions for the WMF community, whose discussions around privacy risks have been documented in community threads.\footnote{See, e.g., \url{https://phabricator.wikimedia.org/T207171}.} Nevertheless, these attacks demonstrate the vulnerabilities of only using heuristic methods such as thresholding (and rounding) to release by-country data and motivate the consideration of more robust protections such as DP.

\subsection{Documentation} 

We designed and developed documentation for each of the three datasets\footnote{All study materials can be found at \url{https://jayshreesarathy.net/wikipedia-study.html}} and presented these to participants before asking them to complete tasks pertaining to the datasets. In this section, we describe the design and iteration process.

\paragraph{Beyond Existing Frameworks.}
Prior work has developed frameworks for communicating data and model information to data users~\cite{gebru_datasheets_2021,mitchellModelCardsModel2019a,crisanInteractiveModelCards2022a}, and disclosing privacy information to data subjects~\cite{kelley2009nutrition,dibia2025we}. 
However, these frameworks do not fully address the challenges of developing documentation for data users about privacy noising mechanisms and the impact on data analyses. An open research direction for usable data privacy is how data curators can effectively convey the impact of privacy noise to data users~\cite{cummings2023centering}--from aligning documentation with users' existing perceptions of uncertainty in data, accounting for users' lack of background in statistics or data privacy, and pinpointing accuracy and utility metrics that are relevant to the data users' particular contexts and tasks. In this study, we focus our documentation on these aspects that are not addressed in prior work. 
We see our work as a starting point to developing effective documentation for privacy-noised datasets.

\paragraph{Design Goals.}
Documentation in this study served as a design probe to understand how participants engage with the privacy-noised data. Thus, our goal was not to develop comprehensive documentation containing all information recommended in prior work, but rather to present basic information about the data and privacy noising mechanism. We aimed for the documentation to be general, rather than directly tailored to the analysis tasks in our study (described in Section~{\ref{subsec:tasks}}).

Using our study team's collective expertise in usable security, privacy communication, data science tools, visualization, DP and Wikipedia data products, we developed an initial set of documentation. We included information about how the data were collected and for what purposes, data processing steps, privacy protections applied, and basic descriptions of data quality and error, organized via the following subsections:
\begin{itemize}
    \item Introduction
    \item Dataset schema
    \item How were the data pre-processed? 
    \item How was privacy protection applied?
    \item How does privacy protection affect the accuracy of the published data?
\end{itemize}
The first three subsections are aligned with prior work, capturing information within the `Motivation,' `Composition', `Collection', and `Pre-processing' categories in the Datasheets for Datasets framework~{\cite{gebru_datasheets_2021}}. The latter two subsections are new to this work, as they detail specifics of the privacy noise mechanisms and how privacy noise impacts the accuracy of the data. 

\begin{figure}[ht!]
\begin{subfigure}{.31\textwidth}
  \fbox{\includegraphics[width=\linewidth, alt={A partial screenshot of the documentation displayed to participants for the DP dataset. This screenshot includes the introduction, data pipeline, dataset schema, and pre-processing steps.}]{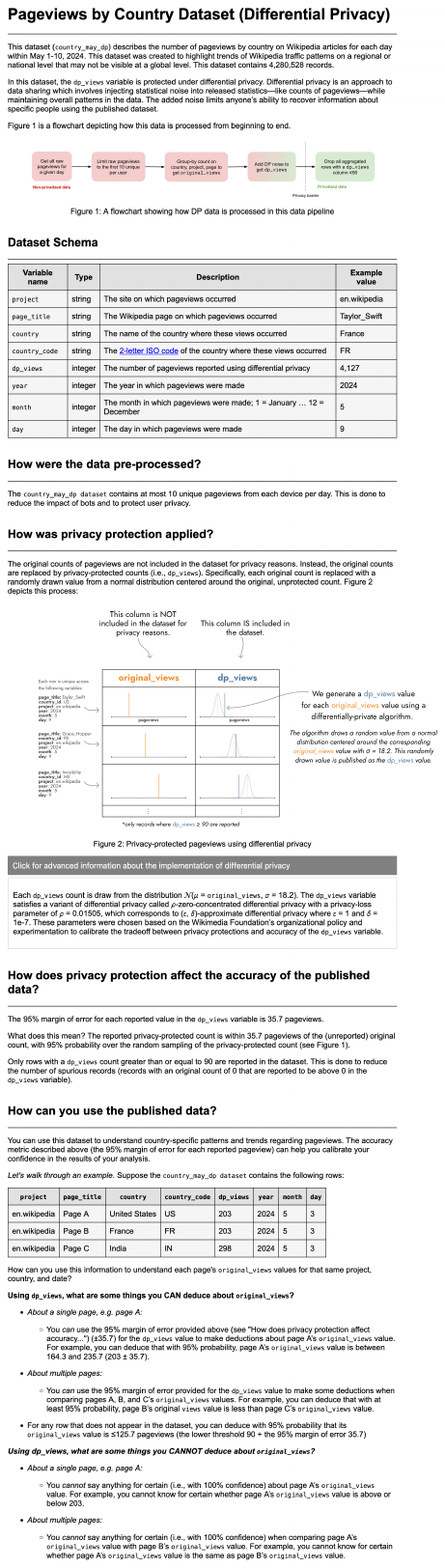}}
    \caption{Introduction, data pipeline, dataset schema, and pre-processing}
    \label{fig:dp-doc-p1}
\end{subfigure}%
\hfill
\begin{subfigure}{.32\textwidth}
  \fbox{\includegraphics[width=\linewidth,alt={A partial screenshot of the documentation displayed to participants for the DP dataset. This screenshot includes the privacy protection mechanism and impact on data accuracy.}]{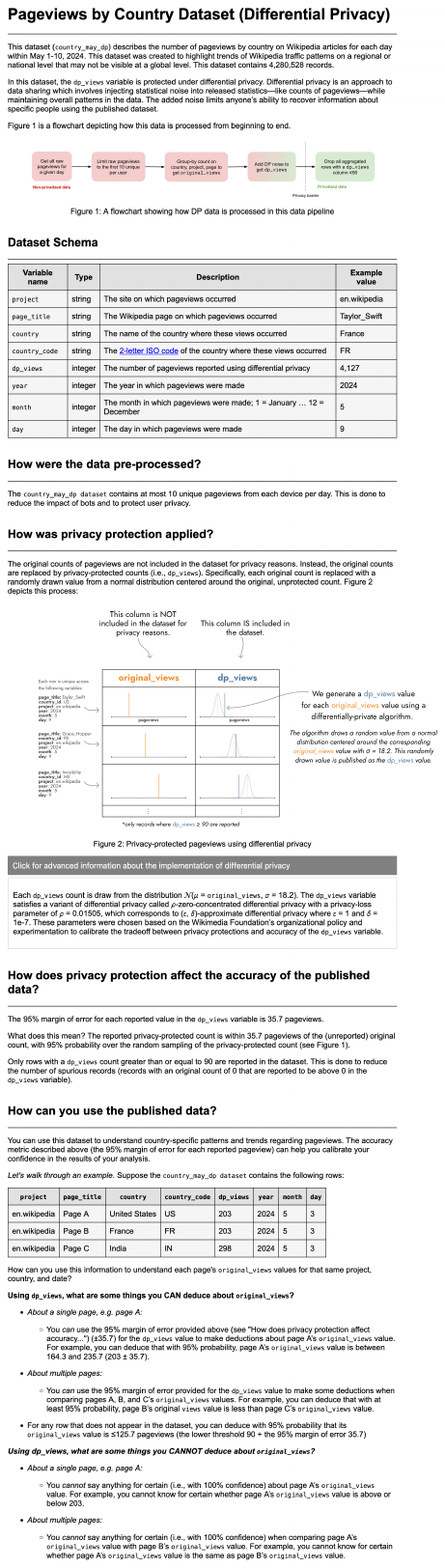}}
    \caption{Privacy protection mechanism and impact on data accuracy}
    \label{fig:dp-doc-p2}
\end{subfigure}
\hfill
\begin{subfigure}{.31\textwidth}
  \fbox{\includegraphics[width=\linewidth, alt={A partial screenshot of the documentation displayed to participants for the DP dataset. This screenshot includes guidance on drawing conclusions using the dataset, with examples.}]{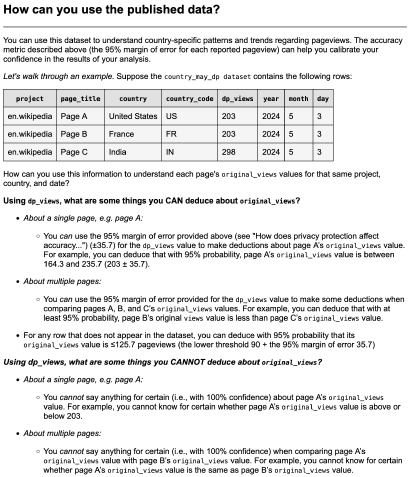}}
    \caption{Guidance on drawing conclusions using the dataset, with examples}
    \label{fig:dp-doc-p3}
\end{subfigure}
\caption{Screenshots of documentation for Pageviews by Country Dataset, where pageviews are reported at a country level using DP.}
\label{fig:dp-docs-p123}
\end{figure}

\begin{figure}[ht!]
\begin{subfigure}{.31\textwidth}
  \fbox{\includegraphics[width=\linewidth, alt={A partial screenshot of the documentation displayed to participants for the Rounded dataset. This screenshot includes the introduction, data pipeline, dataset schema, and pre-processing steps.}]{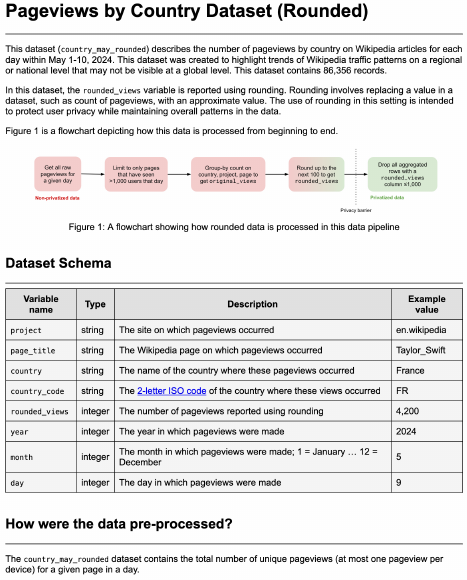}}
    \caption{Introduction, data pipeline, dataset schema, and pre-processing}
    \label{fig:rounding-doc-p1}
\end{subfigure}%
\hfill
\begin{subfigure}{.32\textwidth}
  \fbox{\includegraphics[width=\linewidth, alt={A partial screenshot of the documentation displayed to participants for the Rounded dataset. This screenshot includes the privacy protection mechanism and impact on data accuracy.}]{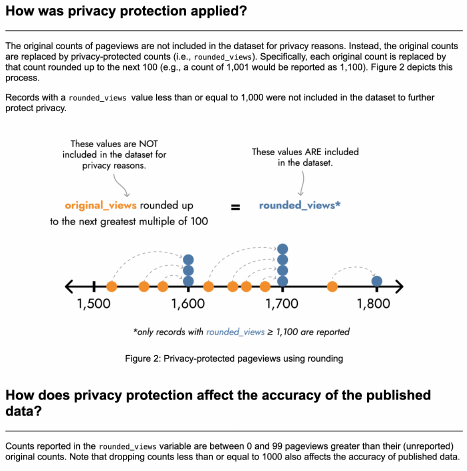}}
    \caption{Privacy protection mechanism and impact on data accuracy}
    \label{fig:rounding-doc-p2}
\end{subfigure}
\hfill
\begin{subfigure}{.31\textwidth}
  \fbox{\includegraphics[width=\linewidth, alt={A partial screenshot of the documentation displayed to participants for the Rounded dataset. This screenshot includes guidance on drawing conclusions using the dataset, with examples.}]{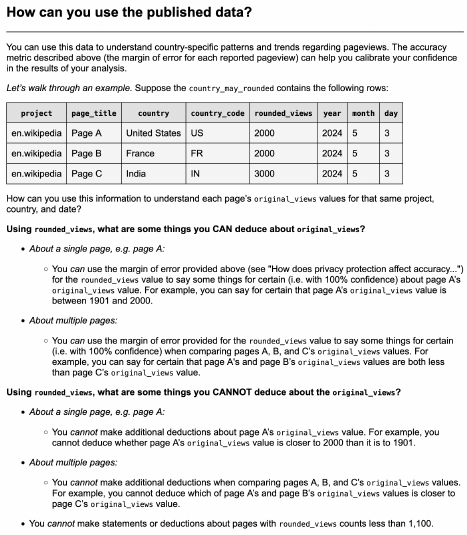}}
    \caption{Guidance on drawing conclusions using the dataset, with examples}
    \label{fig:rounding-doc-p3}
\end{subfigure}
\caption{Screenshots of documentation for Pageviews by Country Dataset, where pageviews are reported at a country level using Rounding.}
\label{fig:rounding-docs-p123}
\end{figure}

In describing the privacy mechanisms, we aimed to preserve the affordances of each technique. For example, we suspected that rounding would be easier to explain but harder to provide certain accuracy metrics, as compared to DP where we could be fully transparent with the implementation details and accuracy implications. We avoided glossing over these differences and presented information naturally in accordance with the capabilities of each method. 
Beyond basic descriptions of the two privacy noising mechanisms, we did not include mathematical definitions of DP or rounding as we did not believe these would be directly relevant for data users.
Instead, we developed original visualizations to describe DP and rounding 
as they were applied by WMF. These visualizations are displayed within the documentation screenshots in Figures~{\ref{fig:dp-doc-p2}} and~{\ref{fig:rounding-doc-p2}}, and reproduced at a larger scale in Figures~{\ref{fig:dp-viz}} and~{\ref{fig:rounding-viz}}.

\paragraph{Expert feedback and iteration}
After developing the initial versions of documentation, we obtained feedback from experts and iteratively updated the documentation. We recruited experts in data privacy and privacy communication using our professional networks and by surveying the space of
privacy-preserving data releases~{\cite{desfontainesListRealworldUses}}. 
We specifically targeted experts engaged in public data releases where privacy has been a concern, who have attended to the communication challenges surrounding these releases. Experts were knowledgeable about DP, as well as heuristic approaches such as rounding, suppression, and anonymization, which allowed them to speak to how both rounding and DP should be presented.\footnote{As the set of such experts is small, we do not report on the demographics or professional backgrounds of experts we interviewed to maintain their privacy.}

We conducted 1-hour semi-structured interviews with five experts. All procedures were approved by our institutional IRBs, and participants were compensated for their time. We asked experts what they consider important information to include in documentation of privacy-noised datasets, standards they are aware of for documenting DP datasets or datasets in general, perceptions on the documentation we had designed for this study, and suggestions for what to include, adapt, or remove from the documentation. 

We provide a summary of findings from expert interviews here; the full protocol and findings are discussed in Appendix~{\ref{app:expert_interviews}}. First, experts were pessimistic about the current state of knowledge around communicating about privacy-preserving mechanisms, both for DP and more broadly for uncertainty in data. Experts agreed that there are no set standards for such communication, but pointed to some resources that start to outline best practices. Experts also emphasized that it is important to design and evaluate documentation in context. With regard to the specific context and documentation we presented, experts had differing views on how much technical information to include in the documentation for Wikipedia data users. Experts hypothesized that the rounding mechanism would be easier to understand but lead data users to be overconfident about what they could do with the data.

\begin{figure}[ht!]
    \centering
    \fbox{\includegraphics[width=.5\linewidth,alt={A screenshot of the documentation displayed to participants for the global dataset.}]{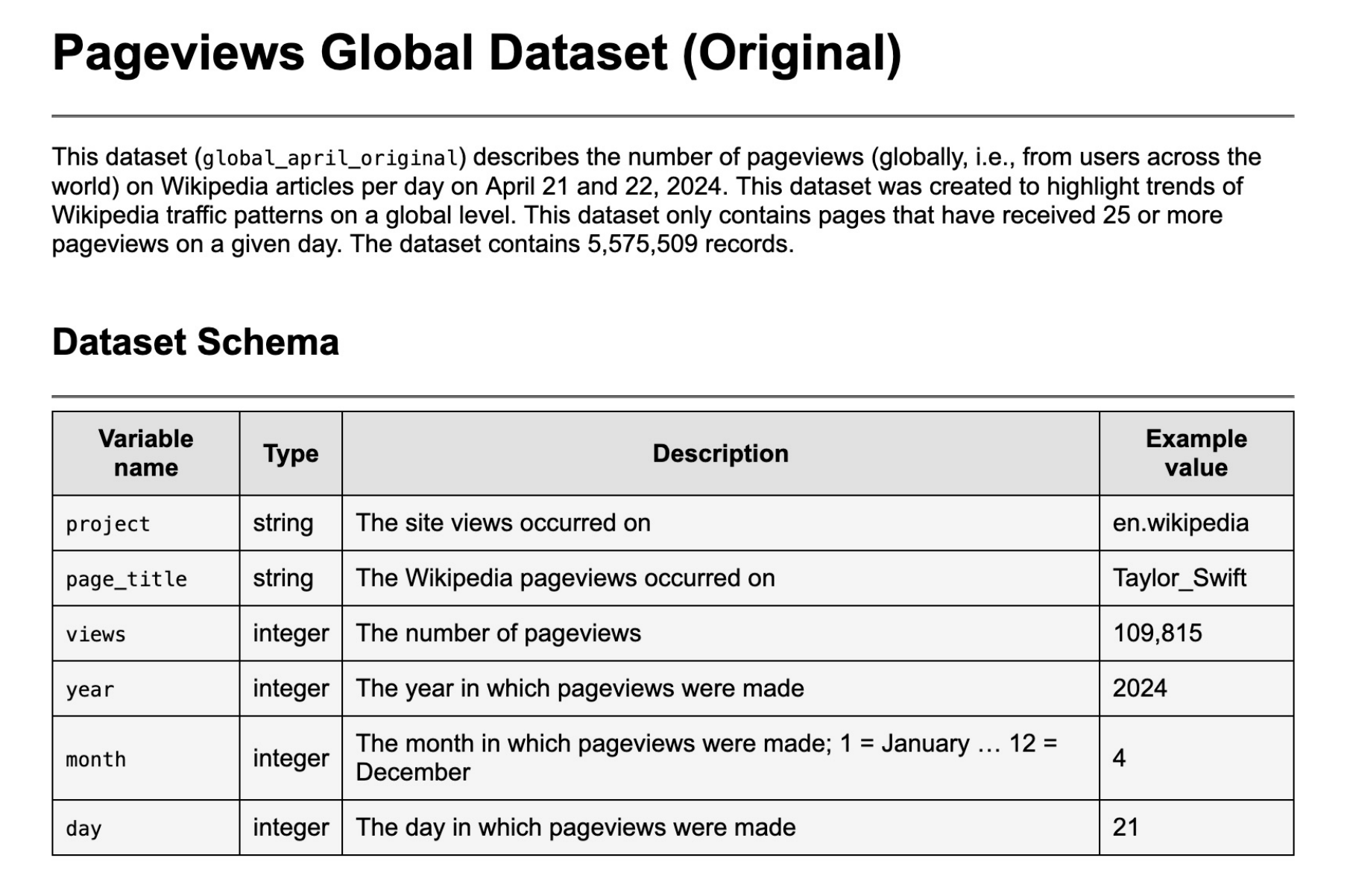}}
    \caption{Documentation for Pageviews Global Dataset, where pageviews are reported at a global level with no privacy protections applied.}
    \label{fig:global-doc-full}
\end{figure}

We made several changes to the documentation based on expert feedback, including: 
adding schematics of the data processing pipelines (Figs~{\ref{fig:dp-doc-p1}} and~{\ref{fig:rounding-doc-p2}}), moving advanced technical information into a expandable box so as to not overwhelm data users (Fig~{\ref{fig:dp-doc-p2}}), and developing a new section titled ``How can you use the published data?'' that provides examples to demonstrate what conclusions can or cannot be safely made using the datasets (Figs~{\ref{fig:dp-doc-p3}} and~{\ref{fig:rounding-doc-p3}}). The updated documentation was rendered on webpages\footnote{See study materials at: \url{https://jayshreesarathy.net/wikipedia-study.html}}, screenshots of which are provided in Figures~{\ref{fig:global-doc-full}}, {\ref{fig:dp-docs-p123}} and~{\ref{fig:rounding-docs-p123}}.

\subsection{Tasks} 
\label{subsec:tasks}

\begin{figure}[t]
\centering
\includegraphics[width=\linewidth,alt={A conceptual representation of the three tasks we asked participants to complete. Task 1 is a maximum value from a set of pages, Task 2 is the average of that page's values across 10 days, Task 3 is an equality check between two pages whose noised values are the same.}]{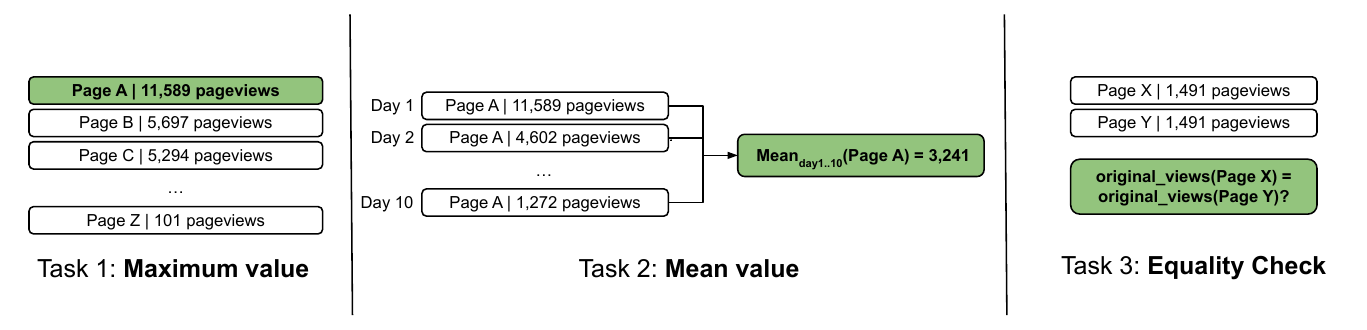}
\caption{Visual representations of the three tasks we asked participants to complete.}
\label{fig:tasks}
\end{figure}

For each dataset, participants were given three data analysis tasks, which are illustrated in Figure~\ref{fig:tasks}. These tasks were chosen to represent common tasks that Wikipedia data users conduct on these datasets. To design tasks, we surveyed papers that used data published by WMF and drew on the expertise of our study team in data science and the given data context. We provide the task templates below; the full tasks are detailed in Appendix~\ref{app:interview_tasks}. In the task templates, \texttt{<VARIABLE>} refers to either the \texttt{rounded\_views} or \texttt{dp\_views} variable.
\begin{enumerate}
    \item[Task 1] Which page had the highest value of \texttt{<VARIABLE>} on \texttt{<DAY>} on \texttt{<LANGUAGE>} Wikipedia in \texttt{<COUNTRY>}? Provide the 95\% confidence interval for the corresponding value of \originalviews~.
    \item[Task 2] For the page with the highest value in Task 1, what was its average daily value for \texttt{<VARIABLE>} across the entire dataset (May 1-10, 2024)? Provide the 95\% confidence interval for the corresponding value of \originalviews~.
    \item[Task 3] On \texttt{<DAY>}, \texttt{<PAGE\_A>} and \texttt{<PAGE\_B>} both have a \texttt{<VARIABLE>} value of $X$ on \texttt{<LANGUAGE>} Wikipedia in \texttt{<COUNTRY>}. Suppose that you're interested in knowing how close \texttt{<PAGE\_A>}'s \originalviews~value is to \texttt{<PAGE\_B>}'s \originalviews~ value. Specifically, imagine that you want to calculate the likelihood that the \originalviews~ values for these two pages are equal.
    \begin{itemize}
        \item Do you think it is possible to complete this task, given the information you have? If so, describe how you might approach this task. You do not need to actually compute anything, unless you would like to.
        \item How would you calculate the likelihood that the \originalviews~ values for these two pages are within 1 pageview of each other?
    \end{itemize}
\end{enumerate}
For each task, we asked participants to describe their thought process however they were comfortable doing so (through code, analytical statements, intuition, etc.). We prompted participants to think aloud and asked follow-up questions as needed to understand their thought processes, particularly around confidence intervals or uncertainty statements.
\section{Methods}
\label{sec:approach}

To answer our research questions, we conducted a task-based analysis study.  Each study session included a think-aloud protocol, contextual inquiry using data analysis tasks, and a semi-structured interview.
Below, we describe our recruitment methods and participant sample, study protocol, and analysis methods.

\subsection{Participants}\label{s.participants}

We recruited participants who were at least 18 years old, familiar with Python or R and Jupyter notebooks, and experienced in data science. 
We aimed to select participants who had worked with data published by WMF, or similar social media and web datasets. 

We posted our recruitment message on LinkedIn and relevant email lists. We also reached out to our professional networks and used snowball sampling to recruit additional participants. 
We determined potential participants' eligibility using a Qualtrics screener survey. At the start of each session, we confirmed participants' eligibility by asking follow-up questions about their experience in data analysis. We discontinued two study sessions because the participants (P6, P10) were not comfortable with data analysis, despite indicating such experience in our screener survey. We did not include data from either of these sessions in our analysis. Overall, we ended up with a total sample of 15 participants for the study, summarized in Table~{\ref{tab:all-participants}}.
Each participant was compensated with a \$50 gift card upon completing the study. 

We asked participants about their background and experience in the screener survey, as well as in the semi-structured post-task interview, and summarized their answers into a label (none, low, moderate, high) as shown in the Table. 9 out of 15 participants had prior experience working with WMF traffic data, and 3 additional participants had experience working with similar web data or downstream data about how people consume Wikipedia information. All participants were experienced in quantitative data science, and 2 participants (P3, P8) additionally worked on qualitative data analysis. Participants were largely inexperienced with DP and rounding: based on our initial screener survey, 8 out of 15 participants had heard of differential privacy, but when we asked them about their prior experience, only 3 participants (P11, P15, P16) had any familiarity with the definition or usage; these participants self-reported at most a moderate understanding. 10 out of 15 participants had passing knowledge of rounding but only 3 participants (P8, P11, P12) had any  understanding of its use for privacy protection; these participants self-reported at most a moderate understanding.

\begin{table}
\begin{center}
\begin{tabular}{|l|l|l|l|}
    \hline
    \textbf{Participant} & \textbf{WMF data experience} & \textbf{DP Familiarity} & \textbf{Rounding Familiarity} \\
    \hline
    P1 & Yes & Low & Low \\ 
    P2 & Yes & Low & Low \\ 
    P3 & Yes & None & Low \\
    P4 & Yes & None & None\\
    P5 & Yes & None & None\\
    \sout{P6} & - & - & -\\
    P7 & Yes & Low & None\\
    P8 & Yes & Moderate & Moderate\\
    P9 & No, but similar & Low & Low \\ 
    \sout{P10} & - & - & -\\
    P11 & Yes & Low & Moderate\\
    P12 & No & None & Low \\
    P13 & No, but similar & None & None \\
    P14 & Yes & None & Low \\
    P15 & No, but similar & Low & Low\\
    P16 & No & Moderate & Low \\
    P17 & No & None & None \\
    \hline
\end{tabular}
\caption{Summary of participant sample for our study with data users. All data users were at least 18 years old and experienced in data science.}
\label{tab:all-participants}
\end{center}
\end{table}

\subsection{Study Protocol}
We briefly summarize our study protocol here, which we refined via two pilot interviews.
First, we asked participants to read through the documentation for the global pageview dataset and think aloud~\cite{willisCognitiveInterviewingPractice2005} about their impressions of the dataset and analyses they would like to conduct with the data. We prompted them to comment on the (un)reliability of the pageview counts in this dataset. Participants were then asked to open the notebook corresponding to this dataset and run cells to view and filter the dataset in a Python or R Google Colab notebook.

Next, participants were asked to read through documentation for one of the two privacy-noised datasets: rounding or DP. The choice of privacy-noised dataset that each person was assigned to interact with first was counterbalanced across the participants. Participants then conducted three analysis tasks on that dataset (see Sections~\ref{subsec:tasks} and~\ref{app:interview_tasks}). 
Participants then followed the same steps—reading documentation, conducting three data analysis tasks, and thinking aloud as they did so—on the other privacy-noised dataset. 

To conclude the study, we conducted a semi-structured interview with each participant (see interview guide in Section~\ref{app:interview_script}) to understand their background and experience with Wikimedia data and privacy noise mechanisms, and perceptions of both privacy-noised datasets. 
We asked participants to reflect for both datasets on analyses that they felt were possible or not possible due to the incorporation of privacy noise, their comfort with and preference for doing data analysis with the datasets, their perception of privacy protections offered by the datasets, and their comfort with  communicating results based on the datasets.

This study was deemed exempt by IRBs at our study team members’ institutions. More ethical considerations about our study protocol are detailed in Appendix~\ref{app:ethical-considerations}.

\subsection{Analysis}
We began to reach saturation after about 12 completed sessions. We conducted an additional 3 sessions to confirm saturation. Overall,
we completed 15 interviews, each of which lasted between 55 and 75 minutes. 
One study team member took detailed notes during each session and cleaned each of the transcripts. Our interviews yielded about 15 hours of recordings and over 200 pages of transcripts and notes. 

We analyzed these recordings, notes, and transcripts using reflexive thematic analysis \cite{braunUsingThematicAnalysis2006}, a standard approach in qualitative HCI studies~\cite{sarathy_dont_2023,song2024inherently}. 
Two study team members familiarized themselves with data from each of the interviews and inductively coded each interview transcript. Throughout the process, we were mindful of our own positionality as researchers; the codes were contextualized not only within the backgrounds of participants, but also within our perceptions of the interviews and our collective expertise in data science, DP, and Wikipedia data. 
Through multiple rounds of discussion, we consolidated the codes and developed themes and findings, which were then presented to the full study team and iteratively refined. The full codebook organized by themes is provided in Appendix~\ref{app:codebook}.

\subsection{Limitations}
The primary methodological tension we faced was creating documentation for non-private data, DP data, and rounded data that were accurate to the details of each dataset and privacy mechanism while still enabling meaningful comparison. The three datasets were each created by WMF as distinct technical artifacts for various purposes; we chose to create a \textit{shared template} for documenting these artifacts, but with differing details between them. This serves as both a strength and a limitation: our study is situated in a real-world context and provides rich qualitative insights, but it is not meant to provide generalizable findings. Finally, the study's explicit focus on privacy and utility evaluation may have influenced participants to comment more on these dimensions of the data than they would have otherwise.

\section{Findings: Data users' engagement with privacy-protected datasets}
\label{sec:findings-rq1}
We present five main findings regarding how data users perceive, interact with, and interpret noise injected for privacy protection as they completed tasks on the Wikipedia datasets. 
Specifically, we discuss participants' baseline understanding of noise in the global (non-privacy-protected) data (Section~\ref{subsec:findings-general-noise}), comfort with understanding the rounding process and with using the DP datasets (Section~\ref{subsec:findings-comfort-DP-rounding}), difficulty computing confidence intervals for multiple perturbed datapoints (Section~\ref{subsec:findings-simple-metrics-documentation}), mixed understandings of both methods' effectiveness at protecting the privacy of data subjects (Section~\ref{subsec:findings-privacy-perceptions}), and contextual preferences about communicating about the rounded versus the DP dataset (Section~\ref{subsec:findings-communicating}). 

\subsection{Participants' baseline understanding of (non-privacy) sources of noise allowed them to think through the impact of additional noise for privacy}
\label{subsec:findings-general-noise}

To contextualize how data users perceive privacy-noised data, we began by exploring their perceptions of a public, non-sensitive dataset that had \textit{not} been privacy-noised. Many participants were familiar with this dataset, which reports pageviews for Wikipedia pages by project and day on a global level (i.e., not broken down by country of reader). When asked to reflect on the reliability of these global pageviews, participants described how they considered these counts to be rough estimates, ``a proxy for utility of a page'' (P12), 
rather than perfectly accurate representations of some true, underlying count of pageviews. 
Participants who had prior experience with Wikipedia data brought up several factors that impact data collection and processing, such as de-duplication of repeated views to the same page, page redirects, impact of bots, translation errors, influence of real-world events, and pages not included in the dataset. 
Several participants described how important it was to account for these factors when conducting and reporting analyses, and how often it came up in their daily work:
\begin{quote}
If you ever get me as a reviewer, that's going to be, like, the first words out of my mouth... ``did you correct for redirects?'' Because it can make a difference up to 25\% in terms of the data. I kick papers out of review all the time for this problem. (P3)
\end{quote}
Other participants were similarly well-versed in the potential sources of error. One participant noted that although they view Wikipedia as a trusted source, they always ``hedge [any] results very carefully'' and only discuss pageviews in ``relative terms'' (P13).

This baseline understanding of pageview data as inherently noisy shaped participants’ views on the impact of intentionally adding noise for privacy. Later in the interview process, when asked about the impact of noise addition on analyses, several participants felt that both DP and rounding would not considerably limit analyses compared to non-private data 
given that the data were already treated as useful for relative rather than absolute statistics: 
even with adding noise, one participant felt the data are ``adequate to perform statistical tests as to whether one page is more popular over time'' (P9).
Another participant tied this idea back to the reliability of the non-private numbers, saying: as ``Internet data is super messy to begin with... messifying it more for privacy would not take away from the kind of work that people are already doing'' (P13). This participant pointed to aggregation itself and other manipulations ``much earlier in the process'' (P13) as steps that could obscure the data more than noise added for privacy reasons. 

When probed about specific tasks, however, participants noted that noise would have varying impacts across analyses: less-viewed pages were more sensitive to noise, while more popular pages were less affected.
\begin{quote}
    The relative impact for, let's say, a page with 1,000 views versus a page with 10,000 views per day will be very different. So if I want to study those very popular pages, then both privacy protection methods we have seen today are not gonna matter a lot, because the base will be just so huge. (P7)
\end{quote}
While participants' existing knowledge of dataset noise made them accepting of added privacy noise, the data analysis exercises clarified how such noise actually affects their analyses.

\subsection{Participants more easily grasped the rounding process, but could devise simulation-based approaches more effectively using the DP-protected data}
\label{subsec:findings-comfort-DP-rounding}

As participants read the documentation, worked through tasks, and reflected on their experiences, we interrogated their comfort with analyzing both the rounding and DP datasets—and how this comfort changed during the course of the study. 
Initially, participants felt that rounding was simpler than DP to understand. As P3 explained: ``The rounding is mathematically easier. It's, like, high school math versus college math.'' Despite perceived simplicity of the rounding noise mechanism, however, participants found that the rounded data provided less useful information to help them assess the uncertainty of their estimates. P16 explained: “Rounding is a lot easier to explain, but it feels more... statistically fraught somehow.''

When thinking through the impacts of rounding on specific analyses, participants were particularly concerned about how rounding could obscure small variations in pageviews. One participant that viewed the rounding data after they worked through the DP data described that ``this sort of next greatest multiple of 100... it's sort of an even larger potential change than the 35.7'' (P3), where the 35.7 refers to the 95\% confidence interval of the DP noise mechanism (see Figure~\ref{fig:dp-viz}). 
The participant felt this scale of noise was a limiting feature of the rounding method:
\begin{quote}
    There's only some research questions that are suitable for a dataset like this... 
    If I'm studying things that are rarely viewed, where a difference in day-to-day views is 1 versus 50, I think that 1 versus 50 is important. Then this dataset is just sort of wholly unsuitable, because that variation is obscured. (P3)
\end{quote}
Once participants started working through the tasks, which prompted them to compute confidence intervals for their analyses, they additionally noted that it was challenging to not know the distribution of noise added to the original values via the rounding process. This forced them to make assumptions about the underlying data distribution.
\begin{quote}
There could be complications in how Wikipedia pageviews are distributed for the rounding task. Like, if they're exponentially distributed, does that mean that I’d have to make inferences about the exponential distribution over an interval...
and then think about, ``Oh, what's the 95\% confidence interval on this chunk of an exponential distribution?'' That would be kind of annoying. I feel like the math with Gaussians is easier to deal with. It's a known distribution...it's more mathematically tractable. (P14)
\end{quote}
The need to make assumptions on the rounded data was especially salient when participants considered simulation methods to estimate confidence intervals. Many participants made an assumption that the underlying data was uniform, which they felt was ``a not-terrible way for us to go about approximating this,'' (P12), but they pointed out that doing simulations to estimate the confidence intervals would only be useful if their assumptions reflected reality. 
Participants commented that while rounding ``makes the numbers nice and pretty'' (P13), it gives you less information about the range of values and thus provided less usability than DP for data analysis. 
\begin{quote}
We just have no way of inferring any distance between the original view to the rounded view, except for within the 100 range... You could probably do another simulation but it would be pretty random, like, it wouldn't really reflect the actual values. (P8)
\end{quote}
For this reason, participants felt that the DP dataset was more suitable than the rounding dataset for running simulations to estimate the impact of noise on their analyses. Many participants
described bootstrapping or simulation-based methods when working through tasks on the DP dataset, whereas only three participants 
mentioned a similar approach for tasks using the rounding dataset. In addition to noting that the distribution of values for DP is well-defined, 
participants highlighted that DP has consistent uncertainty across rows which enables predictable analytics.
Others pointed out that the DP dataset provides unbiased estimates, in that the \dpviews~ value is equal in expectation to the \originalviews~ value, whereas the \roundedviews~ value will always be greater than or equal to its corresponding \originalviews~ value:
\begin{quote}
With differential privacy, the intuition is that things will generally be closer to the true value. With the rounded, your mean will always be too high but with ... DP your mean should be an unbiased estimate of the mean and that feels preferable. (P9)
\end{quote}

Overall, we found that as participants read through the documentation and worked through the tasks, their comfort with the DP dataset increased and their comfort with the rounded dataset decreased. As one participant noted after working through the tasks:
\begin{quote}
I would be able to say useful things about, like, ``Oh, here's this quantity, and here's how certain we are about it,'' and I feel like I would be able to do that more effectively with the differential privacy dataset than the rounded dataset. (P14)
\end{quote}

\subsection{Participants struggled to compute confidence intervals across multiple perturbed data points}
\label{subsec:findings-simple-metrics-documentation}

We noticed that participants were adept at pattern matching with statements from the documentation. For Task 1 (identifying the page with the most reported pageviews and constructing the 95\% confidence interval for that page's \originalviews~ value) 
on both datasets, 14 out of 15 participants were able to connect the question back to what they had read in the documentation and correctly identify the 95\% or 100\% confidence interval for the pages that had the highest DP or rounded views.

However, participants found the calculation of confidence intervals for Task 2 (calculating a page's average reported pageview across 10 days and constructing a 95\% confidence interval for the average \originalviews~ value)
and Task 3 (for two pages that have the same reported pageview, calculating the probability that the \originalviews~ values for these pages are equal) 
challenging. Participants were noticeably self-conscious about not being able to analytically compute confidence intervals, commenting that they felt ``dumb right now'' (P14) and needed to ``brush up on their basic statistics'' (P13). One person reflected: ``There was one point where I was like, `Oh no, it feels like I came to 10th grade stats but I didn't study for the test!' '' (P8). Across all interviews, only a few people were able to outline closed-form solutions for the confidence intervals.

We found that in particular, participants had difficulty applying statistical thinking about confidence intervals when moving from individual rows to aggregation or comparison across rows. For example, participants struggled to make this jump when considering Task 2 
on the DP dataset:
\begin{quote}
I seem to have easily grasped that a DP is a point estimate that comes from a sample, and it basically represents an interval... I can represent that uncertainty at the row level pretty clearly in my mind...
What's confusing for me is what we can say about it getting preciser or less precise. (P13)
\end{quote}

When asked to compare their answers for Task 2 and Task 1, by reasoning about the size of the confidence interval for the average pageview over ten days 
compared to the confidence interval for a single day's pageview, 
many participants displayed incorrect intuition of how noise would aggregate for both privacy methods.
Several theorized that when considering the confidence interval for the average pageview over 10 days for Task 2, ``the inaccuracies would just keep on increasing'' (P4), the confidence interval ``would be getting bigger'' (P8), ``we're introducing more uncertainty'' (P12) and ``adding more noise to the situation'' (P13), and that ``there would be more variance'' from more days of data (P14). As one participant explained, they felt that error would compound like ''interest in a CD account, how you get maybe, like, 3.5\% added, but then it aggregates to its new number, so it gets even bigger'' (P15).

Overall, only three participants (two of whom had recently either taken or taught classes on statistics) correctly identified that for the DP datasets, the confidence interval for the average DP pageview over $10$ days would actually \emph{decrease} by a factor of $\sqrt{10}$, while for the rounded datasets, the confidence interval over those days would \emph{stay constant}. Thus, moving from error on single estimates to confidence intervals across multiple noisy data points seems to be a significant challenge for data users, one that tools and documentation should seek to address, as we discuss in Section~\ref{subsec:effective-use}.

\subsection{Participants had mixed understanding of both methods' effectiveness at protecting privacy}
\label{subsec:findings-privacy-perceptions}

Participants had a range of perceptions about the strength of privacy protections offered by both noising methods. Several participants, especially those who had prior exposure to DP, understood it to provide stricter and more rigorous privacy guarantees than rounding. They attributed these stronger guarantees to various aspects of how the privacy noise is injected. Some pointed to the stochasticity of the process: ``you choose a random number in that range'' (P1) and there’s more ``things in place to kind of protect it... the added noise.'' (P4) Two participants felt that DP was ``a bit more privacy protected because of the possibility of the value being higher or lower'' (P12), and ``you don't know if it's going up or below'' (P15) than the corresponding \originalviews~ value. 

Several of these participants also referenced DP’s status as a ``well-established and well-regarded'' (P9) method in the literature, indicating that their prior exposure to DP may have shaped their views of its protections. However, even some participants who had never heard of DP or used it before pointed to this method as having stronger privacy guarantees because it was ``not deterministic'' (P17).

The perception that DP offered stronger protections was not universal, however. Multiple participants felt rounding to be the safer method \textit{because} of its weaker accuracy. 
For example, one participant said:
\begin{quote}
The DP one feels truer to the data, and so it might make people feel like it's not actually super private because we know kind of the logic of it... With the rounding, with each row I have no idea where within the rounded range it might be. But with DP, I do have some kind of idea of the range. And that's great for data analysis but maybe not for privacy protection. (P8)
\end{quote}
Other participants echoed this line of thinking. One wondered: ``I found the differential privacy more useful, more accessible in terms of making various conclusions. Does that mean that it is less private, though?  I'm not sure'' (P11). Another felt that ``DP is useful in getting a better understanding... but to have an overall view in safer terms, rounding makes sense'' (P5). 
These perspectives were surprising to our study team. In particular, we were struck by how some participants connected a higher perception of data accuracy to a lower perception of privacy, given that this is not always the case. We discuss the implications of this finding further in Section~\ref{shortsec:privacy-utility-tradeoff}.

\subsection{Participants had contextually-dependent preferences for which dataset to use when communicating their results}
\label{subsec:findings-communicating} 
We found that participants expressed preferences over the noise methods based on imagined communication contexts. 
For example, one participant said they would report on rounded data for a non-technical or general audience but use DP data when presenting to an academic research audience, as the standards for precision desired by both audiences would be different in nature (P13). Others similarly felt that their choice of noise methods would depend on the audience's constraints and interests.
\begin{quote}
If I'm talking to a computer scientist, I think I would prefer to use the DP dataset. But I am a social scientist, and some of the people I talk to are a lot more in the humanities, too, and so I think, to avoid going into the math formulas or explaining how the math works there, the rounding one would probably work fine. (P8)
\end{quote}
Participants emphasized that rounding would be ``easier to grasp immediately'' (P8) for audiences that cared about high-level takeaways, and that made it attractive for using in shorter presentations. In these settings, participants felt that audiences would not question the choice of privacy noise mechanism since it is a ``behind the curtain'' (P9) aspect of the data process. 

Participants emphasized that for a technical audience, DP would be preferable because it would be seen as state-of-the-art, ``top of the line'' (P3), and ``good practice'' (P8). 
One participant explained that compared to rounding, the amount of academic literature around DP provides more scientific legitimacy:
\begin{quote}
What you lose with easy explainability you gain with the ability to cite, like, hundreds of papers that ostensibly say that differential privacy is a reasonable thing to do, from a statistical perspective. (P16)
\end{quote}
Another participant called attention to status benefits of using DP, especially when presenting to audiences who have heard of it: ``I think DP sounds sexier... you sound a little smarter if you're using DP instead of rounding'' (P9). 
However, this participant ultimately felt that one’s understanding of the statistical properties was the more important factor in choosing one privacy method over another.

Across many of the interviews, participants were aware of what is signaled by both privacy methods and the numbers they produce. One participant called attention to the aesthetic implications of the rounded numbers: in their view, rounded numbers would signal to an audience that the numbers are noised, whereas DP numbers do not automatically convey the noising process and require explanation:
\begin{quote}
I think from, like, a marketing perspective, it feels like round numbers are more private because they are round...If you're giving a presentation, people will understand that as a heuristic better than, ``Oh, you're giving me a specific number, but it's not... actually the specific number'' 
(P13)
\end{quote}
Overall, participants agreed that DP requires more time and effort to explain and translate to a potential audience. Still, many believed it would be worth doing this extra work in order to present better estimates to the public. One participant said, ``I would feel more comfortable with the differential privacy stuff, just because I feel like if I put in the time, I would be able to back out the math a little bit more effectively, and  be more confident in my estimates of how incorrect my estimates are'' (P14).  Another agreed, saying, ``I guess it's more difficult to communicate the differential privacy approach there. But I get the impression it is the more useful and more valid for the final results actually'' (P11). 

Participants felt that once they explained DP to their audience, the resulting numbers would be more intuitive than the rounded numbers. For some participants, this was because of their feeling that ``most people, even if they're not good at stats, have a good intuition with what a Gaussian means'' (P2) or ``work with a standard interval'' (P17). Another explained that being able to refer to the properties of the normal distribution was helpful and ``makes it feel like that's a more accurate representation of the data than rounding'' (P12).

One participant underscored that doing this translational work of explaining the math of DP would take more time, but that ``people are very smart if you take a minute to walk them through it,'' and ''sticking with the more rigorous approach'' was part of their role and responsibility as a scientist (P3).
This participant emphasized that using the most well-regarded approach was important not just for perceptions, but also in terms of following best practices to minimize risks of data sharing and providing the best results possible. 
\begin{quote}
Well, you're trying to get your work approved through review or what have you, so you kind of want to use the newest, latest, and greatest. But also, I know in my heart that the mathematical properties of the differential privacy stuff has got to be better... and accuracy matters, and measurable accurate degrees of accuracy and and error tracking matters. So I would want to go with the more rigorous technique, not just because it makes me look good, but because that is the standard approach. (P3) 
\end{quote} 
Other participants similarly expressed that it would be important to follow what is regarded as the most rigorous, scientifically-backed approach, both to protect data subjects and to preserve their own reputation as researchers in case the data were to face privacy attacks in the future.

\section{Discussion}
\label{sec:discussion}

Based on our findings from Section~\ref{sec:findings-rq1}, we offer recommendations for data curators and DP researchers to help data users more effectively engage with privacy-noised datasets (Section~\ref{subsec:effective-use}). We describe areas for future research into the implications of noisy numbers for different audiences (Section~\ref{subsec:discussion-context}).

\subsection{Towards effective use of privacy-noised datasets} \label{subsec:effective-use}

\shortsection{Documentation should provide guardrails for what data users can safely deduce from noisy data}
Existing frameworks for documenting datasets recommend including information about whether the data contains confidential or sensitive information, descriptions of pre-processing steps, and guidance on appropriate uses of the data~{\cite{gebru_datasheets_2021}}, but these disclosures do not fully address the impacts of privacy noising mechanisms on the data and analysis process. Documentation of privacy-noised datasets must in addition provide digestible descriptions of the privacy mechanisms applied, how these mechanisms impact data accuracy, and how data analysts can account for the additional noise due to privacy protections in their analysis.

Participants in our study found the ``How can you use the published data?'' section highly informative, and suggested expanding this section with respect to specific analyses--for example, outlining the assumptions an analyst would have to make about the underlying data distribution in order to run simulations and draw statistical conclusions from the noisy data.
Our work provides a clear roadmap for integrating privacy-specific information into existing frameworks (e.g., creating datasheets for privacy-noised datasets)--and provides templates for visualizing the privacy noising mechanisms, providing guidance on conclusions that can or should not be drawn from the data, and explaining the additional information needed by data analysts or steps they must take to account for noise due to privacy. Such guidance can be linked with tooling that helps data users in situ, which we discuss more below.

\shortsection{Describing not only accuracy metrics, but also privacy harms}
Participants questioned the need for privacy protections for by-country pageview datasets. 
Those who had previously used Wikipedia data noted that editor data is regarded as more sensitive than pageview data. 
Others questioned whether it was possible to identify activities from individual users, since the data is already aggregated and thresholded. 
When participants asked us why these protections were made, we provided our default example of how easily some readers could be identified in this data--``imagine a Malaysian speaker in Luxembourg''--and most participants readily understood the possibility for privacy harms for minority population, even without us sketching possible attacks as in Section~{\ref{subsec:datasets}}.

In the documentation we designed, we did not discuss the privacy harms of releasing the by-country datasets without noise, as this information is not directly needed for data users to be able to use the data. However, participants' reactions suggested that data users may benefit from documentation that addresses not only accuracy implications of the noising techniques, but also privacy dimensions of the data, especially for datasets that may not at first blush seem to report sensitive information. This is a new finding challenging the view of some of the experts we interviewed that data users may not care about protecting privacy of data subjects beyond what is legally required, and complements insights from prior work on designing explanations of DP for data subjects~\cite{nanayakkara_what_2023,smart2024models}, which find that data subjects care not only about privacy guarantees but also about the accuracy implications and usefulness of the data. Thus, presenting privacy and accuracy \emph{together} seems to be important when communicating to both data subjects and data users.

\shortsection{Communicating the privacy-utility tradeoff of noisy statistics} \label{shortsec:privacy-utility-tradeoff}
In our study, some data users had surprising perceptions of the relationship between privacy and accuracy of statistics: several participants who had less exposure to the technical privacy literature connected the higher accuracy of DP data to a lower perception of privacy protections offered by DP data (as described in Section~\ref{subsec:findings-privacy-perceptions}). 
This reasoning is incorrect, yet informs us about a potential starting point of data users who must grapple with tradeoffs between accuracy and privacy. 

A core challenge of DP communication has been to convey that there even exists a \emph{privacy-accuracy tradeoff}--informally, 
high accuracy for many statistics implies weak privacy protections for data subjects \cite{dinurRevealingInformationPreserving2003}. Our work suggests that participants \emph{do} consider privacy and accuracy together but may come to incorrect conclusions about the relationship between the two (in particular, it is not true that lower accuracy, such as with rounding, implies stronger privacy protections). Communicating this tradeoff requires first understanding data users' existing understanding of how privacy and accuracy impact one another in the context of noisy data. Future work should explore these prior beliefs in order to design communication materials that effectively convey the privacy-utility tradeoff of statistical releases.

\shortsection{Tooling to help generate confidence intervals} \label{shortsec:tooling}
Although the documentation we provided contained high-level descriptions of the privacy noise mechanisms, visualizations of how the mechanisms were applied, and basic accuracy metrics, participants were unsure how to compute the uncertainty of downstream statistics on noisy data.
Even when participants were experts in data science, many struggled to apply concepts from the documentation to make inferences beyond what was directly stated—for example, when asked to consider confidence intervals across multiple perturbed data points (Section~\ref{subsec:findings-simple-metrics-documentation}). 

To help data users make the jump from accuracy metrics stated about individual data points to accuracy metrics for more complex analyses, we recommend building software tools that can keep the information about the noise mechanism ``attached'' to each step of the analyses. For example, for Task 2 on the DP dataset (calculating the average reported pageview for a single page over 10 days, and constructing a 95\% confidence interval for the average \originalviews~ value for those 10 days), the tool could keep track of the noise distributions of the pageview for each day and provide a confidence interval (using basic statistical properties of i.i.d. Gaussians) for the confidence interval of the mean. We can think of this tool as both extending the life of the documentation (by preserving uncertainty information about each data point in the downstream analysis pipeline) and helping data users leverage closed-form, analytic characterizations of the noise distribution of a resulting statistic when it exists. Several accounting tools exist \cite{pipelinedp, opendp} for composing noise distributions across computations~\cite{GuesstimateSpreadsheetUncertain}, including across various DP mechanisms~\cite{pipelinedp,opendp}. It would be worthwhile to build upon these tools to calculate and keep track of confidence intervals for privacy-noised data throughout the downstream analysis process.

\subsection{What do noisy numbers convey to audiences?}
\label{subsec:discussion-context}
Data are made and interpreted contextually \cite{boydCriticalQuestionsBig2012,gitelman2013raw}. Numbers represent more than a count; they also signal to an audience about the process of creating data and the reliability of the numbers. Yet, these signals may not align with the underlying guarantees of the data. For example, prior work has highlighted that population counts in census data that are reported down to the individual unit signal to data users that the counts are more precise than they actually are~\cite{boyd_differential_2022}.

In this study, participants described how they would consider the implicit signaling of presenting analyses performed on the rounded versus DP datasets when giving a scientific talk. In particular, they felt that using rounded numbers immediately conveys to an audience that the numbers are not precise and have some privacy protections applied, whereas presenting DP numbers requires an additional level of explanation to convey that the numbers have been privatized and are less precise than they may appear. Participants felt that this implicit signaling for each of the methods creates benefits and challenges for communication. 

Recent work by Hod et al.~\cite{hodDifferentiallyPrivateRelease2025} displayed a similar finding when designing a privatized data release, highlighting how stakeholders did not feel that the published DP counts adequately conveyed the strength of privacy protections. The authors decided to post-process the privacy-protected data to remove unique records, in order to align with stakeholders’ understandings of what privacy-preserved data \emph{should} look like, even though this did not affect the privacy guarantees of the data. The authors called this ``face privacy,'' inspired the concept of ``face validity''~\cite{weiner2010corsini}, to describe how private the data \emph{appeared subjectively} to stakeholders aside from the technical privacy protections. 
Our work adds additional examples of what data users might consider to `appear private' in the Wikimedia context, and highlights the need to develop broader theories of public perceptions of privacy protections based on the presentation of numbers or statistics.

Future work may be able to draw on work from history and sociology of statistics. For example, Desrosi\'eres categorizes different statistical communities in terms of how they perceive the ``reality'' of numbers, ranging from business accounting to inferential statistics \cite{desrosieresHowRealAre2001}; these contextual insights can be expanded towards understanding how communities perceive statistical uncertainty for privacy protections. More recently, scholars have described competing `statistical imaginaries' that can lend insight into how statistical methods are perceived by data users~\cite{sarathy2025statistical}. 
Such efforts to understand and contextualize perceptions of noisy data are critical for guiding the design of documentation tailored to specific data user communities.

\section{Conclusion}

Organizations seeking to publish data are encouraged to use noising techniques to protect the privacy of data subjects. However, prior releases of privacy-noised data have resulted in negative reactions from data users, illustrating the challenges of communicating effectively to data users. Without a better understanding of these communities' perceptions of noisy data and resources that support effective data utilization, organizations may shy away from potential pushback and opt to not release data at all. Thus, these gaps threaten the broader project of safe, open data.

This work takes a step toward understanding data users' engagement with noisy data.
We conduct a task-based contextual inquiry and semi-structured interviews with 15 participants, exploring how data users interact with two real, privacy-noised Wikipedia datasets. We design expert-informed documentation for the datasets to ground the study in a real-world analysis environment. 
We find that data users are better able to devise simulation methods to compute uncertainty with DP-noised data than with rounded data, yet for both datasets they struggle to compute confidence intervals across multiple noisy datapoints. Participants have mixed perceptions of privacy protections offered by DP and rounding, shaped by their exposure to DP and their perceived utility of the datasets. Future work should explore the aesthetic and contextual implications of noisy numbers when presented to different audiences. We offer suggestions for designing documentation and tools to help data users more effectively use privacy-noised data.

\section*{Acknowledgements}
We thank  danah boyd, Matt Franchi, Rishi Jha, Isaac Johnson, Jae June Lee, Tom Ristenpart, and Lu Xian for helpful discussions during various stages of this project, Sarah Radway and Ryan Steed for their feedback on visuals within the documentation, and Margaret Haughney for administrative support. We are grateful to the anonymous reviewers for their thoughtful suggestions and to all the participants in our study for making this work possible.

\printbibliography
\clearpage
\appendix
\section{Interviews with DP Experts}
\label{app:expert_interviews}
We conducted 1-hour semi-structured interviews over Zoom with five experts. These interviews were recorded and transcribed with participants’ consent, and we compensated participants with a \$50 giftcard. All interview procedures were approved by our institutions’ IRBs.
Experts had significant experience with DP in practical settings, ranging from 5 to 12 years (average of 8.2 years), across academia, government, nonprofit organizations, and industry.

\subsection{Semi-Structured Interview Protocol} 
\label{app:expert_questions}
We followed the interview guide below when interviewing DP experts, asking follow-up questions where relevant.

\begin{itemize}
\item Imagine you recently released a tabular dataset protected under DP. The dataset will be used by data analysts with some statistical and/or data science background, but no DP background. You want to create documentation that helps them more effectively analyze the dataset, given the additional source of statistical noise introduced by DP. Please talk through what information you think is important to include in the documentation.
\begin{itemize} 
    \item What parts of the documentation do you believe are most salient for data users who are asked to do basic data analysis tasks (sums, means, etc.)? 
    \item What parts of the documentation are important for providing context on uncertainty and calibrating trust in the results?
    \item What, if any, misconceptions or misuses of DP data do you believe are important to address via documentation?
\end{itemize}
\item In your view, how should DP documentation be similar to or different from documentation of datasets with other privacy protections (such as rounding)?
\item Are there any standards for documenting DP datasets that you are aware of? Do you find these standards useful? Are there any standards for documenting datasets in general that you believe are portable to DP settings?
\item \textit{Show the initial documentation we designed:}
\begin{itemize}
    \item Which parts of this documentation do you think data analysts would find useful and why?
    \item Which parts of this documentation do you think data analysts would \textbf{not} find useful and why?
    \item What, if anything, is missing from this documentation?
    \item What pieces of information, if any, do you think we should \textbf{not} include in the documentation and why?
\end{itemize}
\item \textit{Describe our aims for the study and the specific tasks we will ask participants to conduct.}
\begin{itemize}
    \item How well do you think this documentation supports our goal of exploring data users' engagement with noisy data?
    \item How well does this documentation support data users when doing the given tasks?
\end{itemize}
\end{itemize}

\subsection{Findings from DP Expert Interviews}

\paragraph{Experts were pessimistic about the current state of knowledge around communicating about privacy-preserving mechanisms.} One expert said: ``we should assume that we’re doing a bad job of it right now and that we really don’t know what is the best way to do it'' (E1). Specifically for DP, experts pointed to the different definitions, models, and parameters that specify the exact meaning of DP, as a barrier for communication. 
\begin{quote}
    The big issue is that there are different relaxations and approaches to preserving privacy. We’re terrible at communicating it. Local DP is a whole different framework, for example. If you’re deep in the field, it makes sense, but from the outside it’s incredibly messy and complicated. Having multiple tuning parameters is very hard for people to grasp. (E2)
\end{quote}
Multiple experts pointed out that communicating about uncertainty, even general uncertainty that did not pertain to privacy noise, is a difficult problem. One expert had searched for answers across other fields such as statistics and medicine, but felt that there was no great way to communicate or understand uncertainty in data in general.
\begin{quote}
    Even people with backgrounds in statistics don’t have a great framework to understand uncertainty in data. How do you tell them what that noise is like? It’s not clear that there’s a good language to do things like that at the moment. (E1)
\end{quote}

\paragraph{All experts stated that there are no concrete standards or consensus on communicating about privacy-preserving datasets, but they pointed to some resources that start to address best practices.} These included the Epsilon Registry \cite{dwork_differential_2019} and a oft-used list of real world deployments and conversions across privacy parameters \cite{khavkin2025differential, desfontainesListRealworldUses}. One expert also highlighted the recent NIST guidelines on communicating DP~\cite{NISTDP2025}, which they felt was an important starting point:
\begin{quote}
    Having a first attempt out there is better than having nothing out there. Even if it’s not an official standard, a document put out by NIST is still looked at as quite authoritative even if it’s squishy in many ways...we haven’t as a community built the standards, but even if we did, we wouldn’t know exactly what to put in them in every case. (E1)
\end{quote}
Experts mentioned that approaches to documenting data or models in other contexts, such as model cards~\cite{mitchellModelCardsModel2019a} or Croissant \cite{akhtarCroissantMetadataFormat2024a}, could be a helpful starting point for documentation on privacy-preserving datasets. For example, E1 praised the widespread use of model cards for creating ``social pressure'' to release documentation and ``encouraging structured transparency'' (E1), expressing hope that the privacy community can adapt the spirit of a common format and expectation of transparency going forward. But how to do so concretely was less clear. Two experts (E1, E4) emphasized that existing approaches do not fully address the challenges of communicating about privacy noise and cannot be directly adapted for DP. 

\paragraph{Experts emphasized that the context and audience matters for choosing what to include in documentation about privacy noise.} E2, E4 and E5 questioned whether universal standards were the right approach given the different possible data users and use cases, and instead encouraged those creating documentation to really understand their audience. One expert suggesting that designers ask themselves the following questions: ``Who's your audience? What’s the story you’re telling the person? What’s the hook? What are the tradeoffs? What is the key takeaway(s) that you want the audience to know?'' (E2) 

Experts emphasized that it was important to design and evaluate documentation in context.
\begin{quote}
    It’s really use-case dependent. It’s important to talk about each audience’s specific case, rather than generalities and to take the next step into ``here’s how the DP implementation affects your specific use cases.'' (E4)
\end{quote}
One expert noted that in the US Census Bureau’s differentially private data release, the type of communication that landed well depended on the technical level of expertise (E3). This expert explained that data users inside the Bureau were highly sophisticated, so they benefited from having all the information available. But for external data users, it was more effective to tell users how the error impacted their work---for example, that noise is more impactful for small geographies but is less significant for larger geographies.
To be able to understand the needs of data users, this expert (E3) recommended empowering trusted intermediaries and using feedback loops with community liaisons to design the documentation that accompanies data releases. 

\paragraph{Experts had differing views on how much technical information to include in our specific documentation.} Multiple experts felt that DP documentation should include information about the noise distribution and privacy parameters in order to enable privacy experts to vet the privacy protections, but all of them worried that such information could overwhelm non-experts in privacy who simply wanted to use the data. Some experts appreciated that our documentation was careful to not overwhelm non-technical users, saying: ``I personally like this a lot more than most of the documentation that I’ve seen that try to explain DP'' (E1), when explaining the 95\% error metric for the DP counts, the `` `What does this mean?' language is helpful and clear'' (E3) and that overall, the ``concise documentation is good for first-time users'' (E5).

One expert felt that since data users don’t care about the implementation of DP, this technical information should exist in a separate document. However, another expert believed strongly that privacy implementation information was important to include for completeness. Thus, we followed a middle road, an ``accordion approach'' recommended by one expert: we enabled data users to click an expandable box to find advanced information about privacy parameters of DP.
\begin{quote}
    Some people might be intimidated by math and Greek letters. These users just need to know that [the data] is protected but they don’t care exactly how, so you can let them click in to find more information about the privacy parameters. (E2)
\end{quote}

\paragraph{Experts felt that the rounding mechanism was easier to understand, but it may cause data users to be overconfident about what they could do with the data.}
For the rounding documentation, experts were mostly satisfied by how we presented the information pertaining to rounding, but they were worried that data users might make incorrect inferences based on the data. 
\begin{quote}
    I’m worried that people would read this and be like ``Great, this is deterministic, theres no noise'' and then just use it as if there’s no noise. I worry that you could end up making worse decisions than with the DP data. (E1)
\end{quote}
They clarified that this worry was not specific to our documentation but rather pertained to the concept of rounding and how familiar people are with it from an early age. 
\begin{quote}
[Rounding is] something that we teach in grade school, and so you don't need as many explanations... And then, as a user, you're like, ``OK, totally got it. You're doing some privacy protections. It's rounded to nearest 100.'' That's like a WAG for me... a wild-ass guess. So sometimes you just need a WAG, or a SWAG. A SWAG is a scientific wild-ass guess. (E2)
\end{quote}

\subsection{Main changes to documentation based on experts' feedback}
The first main change we made to our documentation based on expert feedback was including a section on `How can you use the published data?' Our initial documentation included basic accuracy statements but not much guidance of how data users could use the given metrics to make conclusions about the data. Multiple experts recommended including such guidance for both the rounding and DP datasets. As one expert explained:
\begin{quote}
    If I’m a data user, I can look at all the documentation and understand what the processing and steps look like. Then I get to the bottom and want to know `Can I be confident in terms of the answer to my question when I’m using this data?’ I now understand the variance, but still unclear what to actually do. What can I actually do to make myself more or less confident in my answer? (E1)
\end{quote}
We added a section of this nature to documentation for both the rounding and DP datasets after the first three interviews. The last two experts we interviewed reacted positively to this section. For example, one appreciated the examples we provided, and said ``this is good because it is talking to the data user much more than the previous sections'' (E4). The other expert said they liked our statements such as ``You cannot say anything about comparisons within the same range'' (E5).

The second main change we made to our documentation was including a data pipeline diagram to explain how the data was processed. This was recommended by E4, who felt that the documentation was missing a process diagram, for example describing the input data, how data was limited by user or by views, how the data was grouped, and how individual rows were dropped based on the threshold for each dataset. As this expert explained, 
\begin{quote}
    This may be from a computer science perspective who thinks about things as an algorithm... I’m not sure about how a statistician would want to see this. But I’d like to see: here are the three things that have been done as part of the process via a diagram. (E4)
\end{quote}
Finally, we made some minor changes to naming of variables:
changing \texttt{privacy\_protected\_views} to \texttt{dp\_views} and \texttt{rounded\_views}. This was based on a comment by one expert that objected to rounding being called an adequate privacy protection, and also to differentiate between the two types of noising mechanisms for this variable.

\section{Task-Based Study: Additional Information}

In this section, we provide additional information about our task-based study with data users, including visualizations of rounding and DP mechanisms, tasks that participants were asked to complete, the full interview guide we followed for each study session, ethical considerations of the study, and the codebook we developed when analyzing the recordings, notes, and transcripts from the study sessions.

\subsection{Visualizations of DP and Rounding}

\begin{figure}[ht]
    \centering
    \includegraphics[width=0.85\linewidth,alt={A visualization of the differential privacy noise mechanism that was presented to study participants. On the left there are original views, represented as an orange vertical line. On the right there are DP views, represented by a Gaussian distribution + a blue line representing the noised value.}]{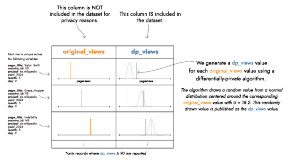}
    \caption{Visualization included in our documentation of the DP process used by WMF to report Wikipedia pageviews by country.}
    \label{fig:dp-viz}
\end{figure}

\begin{figure}[ht]
    \centering
    \includegraphics[width=0.65\linewidth,alt={A visualization of the rounding noise mechanism that was presented to study participants. There's a number line with orange dots along it, representing the original views. Each dot is rounded up by pushing it to the next 100, where it is stacked up and turned blue to represent rounded views.}]{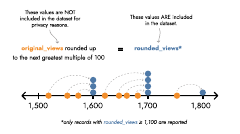}
    \caption{Visualization included in our documentation of the rounding process used by WMF to report Wikipedia pageviews by country.}
    \label{fig:rounding-viz}
\end{figure}

Below, we display the visualizations we developed for DP (Fig{~\ref{fig:dp-viz}}) and Rounding (Fig{~\ref{fig:rounding-viz}}) as they were applied by WMF for the Wikipedia Pageviews by Country Datasets. These visualizations were included in the documentation provided to participants (see Figures~{\ref{fig:dp-doc-p2}} and~{\ref{fig:rounding-doc-p2}}).

\subsection{Data Science Tasks} \label{app:interview_tasks}
Below, we list the data analysis tasks provided to participants in our user study. 
Section~\ref{app:interview_tasks:dp} lists the three tasks participants were asked to complete on the DP dataset, and Section~\ref{app:interview_tasks:rounding} lists the three similar tasks participants were asked to complete on the Rounded dataset. 
These tasks were provided to participants in Colab notebooks, either in Python or R based on their preference, along with baseline template code. These notebooks and other study materials are provided at the following link: 
\url{https://jayshreesarathy.net/wikipedia-study.html}.

\subsubsection{DP Tasks} \label{app:interview_tasks:dp}

\begin{enumerate}
    \label{app:interview_tasks:dp:1} \item[Task 1:] Which page had the highest value of \texttt{dp\_views} on May 1, 2024 on Spanish Wikipedia in Mexico?
    
    Provide a 95\% confidence interval for the corresponding value of \texttt{original\_views}.
    \label{app:interview_tasks:dp:2} \item[Task 2:] For the page with the highest \texttt{dp\_views} value in Task 1, what was its average daily value for \texttt{dp\_views} across the entire dataset (May 1-10, 2024)?

    Provide a 95\% confidence interval for the corresponding average daily value for \texttt{original\_views}.
    \label{app:interview_tasks:dp:3} \item[Task 3:] On May 3, 2024, ``Taylor\_Swift'' and ``Ella\_Purnell'' both had a \texttt{dp\_views} value of 1,474 on English Wikipedia in Canada. Suppose that you're interested in knowing how close Taylor Swift's \texttt{original\_views} value is to Ella Purnell's \texttt{original\_views} value. Specifically, imagine you want to calculate the likelihood that the \texttt{original\_views} values for these two pages are equal.
    \begin{itemize}
        \item Do you think it is possible to complete this task, given the information you have? If so, describe how you might approach this task. You do not need to actually compute anything, unless you would like to.
        \item How would you calculate the likelihood that the \texttt{original\_views} values for these two pages are within 1 pageview of each other?
    \end{itemize}
\end{enumerate}

\subsubsection{Rounding Tasks} \label{app:interview_tasks:rounding}

\begin{enumerate}
    \label{app:interview_tasks:rounding:1} \item[Task 1:] Which page had the highest value of \texttt{rounded\_views} on May 1, 2024 on English Wikipedia in Australia?
    
    Provide a 95\% confidence interval for the corresponding value of \texttt{original\_views}.
    \label{app:interview_tasks:rounding:2} \item[Task 2:] For the page with the highest \texttt{rounded} value in Task 1, what was its average daily value for \texttt{rounded\_views} across the entire dataset (May 1-10, 2024)?

    Provide a 95\% confidence interval for the corresponding average daily value for \texttt{original\_views}.
    \label{app:interview_tasks:rounding:3} \item[Task 3:] On May 9, 2024, ``Los\_Tigres\_del\_Norte'' and ``Justin\_Bieber'' both have a \texttt{rounded\_views} value of 1,500 on Spanish Wikipedia in Mexico. Suppose that you're interested in knowing how close Los Tigres del Norte's \texttt{original\_views} value is to Justin Beiber's \texttt{original\_views} value. Specifically, imagine you want to calculate the likelihood that the \texttt{original\_views} values for these two pages are equal.
    \begin{itemize}
        \item Do you think it is possible to complete this task, given the information you have? If so, describe how you might approach this task. You do not need to actually compute anything, unless you would like to.
        \item How would you calculate the likelihood that the \texttt{original\_views} values for these two pages are within 1 pageview of each other?
    \end{itemize}
\end{enumerate}

\subsection{Study Protocol} 
\label{app:interview_script}

We followed the guide below when conducting our study with data users, asking follow-up questions where appropriate.
Note that we alternated the order in which we showed the DP documentation and tasks versus rounding documentation and tasks to participants. 

\begin{itemize}
    \item In today’s session we will ask you to answer some questions based on three different datasets about pageviews on Wikipedia. Afterward, we’ll ask you some questions about your experience. To begin, we’ll start with the Pageviews Global Dataset.
    \item \textit{Share screen and provide remote control.}
    \item Open up the ``global documentation'' tab. Take a minute to skim this document. As you are reading, please think aloud. It’s very helpful for us to hear your thought process. You may ask questions at any point.
    \item \textit{Allow participant to read global documentation (Figure~\ref{fig:global-doc-full}).}
    \begin{itemize}
        \item What are your first impressions about this dataset? For example, what kinds of analyses might you want to do with this, and what challenges might you face?
        \item How reliable or unreliable do you think these reported counts are? How would you communicate about these counts on this dataset with a colleague?
    \end{itemize}
    \item Now, we’ll have you look through the first notebook to see an example of how to access and filter the dataset. Click over to the ``Global\_tasks'' tab and follow the instructions there.
    \item \textit{Allow participant to follow notebook instructions. There are no problems to solve, they just see examples here.}
    \item We’ll now start the next section of the study, where you’ll be looking at a dataset with pageviews broken down by country. To protect the privacy of Wikipedia readers, this dataset has been protected using differential privacy. It’s OK if you haven’t heard of differential privacy before; it will be explained in the dataset’s documentation. Please navigate to the ``DP dataset documentation'' tab. We'll ask you to complete some data analysis tasks using this dataset. But first, please take a minute or two to skim the document. You'll be able to refer back to this document as you complete the tasks, so don’t worry about getting all the information now. As you skim, please think aloud. You may ask questions at any point.
    \item \textit{Allow participant to read DP documentation (Figure~\ref{fig:dp-docs-p123}).}
    \item Please click on the tab ``DP\_tasks.'' There are three tasks in the notebook. We'll have you complete them one by one, starting with the one at the top. Please think aloud as you read and work through the task. We’re not testing your coding ability; we care about hearing your thought process. Remember that you can refer back to the documentation as much as you want. We have several tasks to get through, and to make sure you finish on time, we’ll be keeping track of time. We’ll give you a heads up when you should be wrapping up. Don’t worry if you don’t finish the task; we’re more interested in your thought process.
    \item \textit{Allow participant to follow notebook instructions and complete DP tasks (Appendix~\ref{app:interview_tasks:dp}).}
    \item We’ll now move on to the next section of the study, where you’ll again be analyzing a dataset with pageviews broken down by country. This dataset was not protected using differential privacy. Rather, it was protected using a rounding methodology. The details of how rounding was applied will be explained in the dataset’s documentation. Please navigate to the ``Rounding dataset documentation'' tab. You will complete some data analysis tasks using the new dataset, but first take a minute or two to skim the document. As you skim, please think aloud. Again, you may ask questions at any point.
    \item \textit{Allow participant to read rounding documentation (Figure~\ref{fig:rounding-docs-p123}).}
    \item Please click on the tab ``Rounding\_tasks.'' There are three tasks here. Again, you’ll complete them one by one. Please think aloud as you read and work through the task. Remember that you can refer back to the documentation as much as you want.
    \item \textit{Allow participant to follow notebook instructions and complete rounding tasks (Appendix~\ref{app:interview_tasks:rounding}).}
    \item \textit{Ask wrap-up interview questions:}
        \begin{itemize}
        \item Can you describe your prior experience with analyzing data published by the Wikimedia Foundation?
        \item Can you describe your experience with privacy protections methodologies (including, but not limited to differential privacy, k-anonymity, rounding, suppression)?
        \item How do you feel about the DP dataset, in terms of the types of analyses you might want to conduct? How do you feel about the rounding dataset, in terms of the types of analyses you might want to conduct?
        \begin{itemize}
            \item Are there any analyses that the privacy protections added to each of the datasets prevents you from conducting?
        \end{itemize}
        \item How do you feel about the privacy protections offered by the both datasets, and why? 
        \begin{itemize}
            \item Follow-up: do you feel like there should be a right answer to this?
        \end{itemize}
        \item Do you feel more comfortable analyzing one dataset over another? If so, why?
        \begin{itemize}
            \item Prompt: Does uncertainty come to mind?
            \item Follow-up: do you feel like there should be a right answer to this?
        \end{itemize}
        \item Do you feel more comfortable reporting analyses to a broad scientific audience from one dataset over another? How about to the general public? If so, why?
        \begin{itemize}
            \item Prompt: Does credibility come to mind?
        \end{itemize}
        \item Are there any other thoughts you’d like to share?
    \end{itemize}
\end{itemize}

\subsection{Ethical Considerations} \label{app:ethical-considerations}
This study was deemed exempt by IRBs at our study team members’ institutions. We were careful to follow best practices throughout the study. During recruitment, we aimed to minimize data collection in our screener survey and followed best practices for snowball sampling (asking contacts to forward our screening survey, rather than providing us with potential participants’ contact information) in order to reduce risk of coercion or privacy harms. Prior to the study, we obtained written consent from each participant. At the beginning of the study, we provided them opportunity to ask questions about the study and obtained verbal consent before recording the session.

The study sessions were conducted by giving participants only subsets of publicly available data from WMF. The global dataset was deemed to not be privacy-sensitive by WMF, and the by-country datasets were published using privacy protections of rounding and differential privacy. Our tasks were not designed to re-identify data subjects in any of these datasets.

Recordings, transcripts, and notes were stored in a secure drive with access limited to study team members that had been approved by their institutions' IRBs to access the identified data. We removed direct participant identifiers before analyzing the notes and transcripts from the interviews.

\subsection{Codebook} \label{app:codebook}

Through multiple rounds of discussion, members of the study team developed codes and themes, following a reflexive thematic analysis approach~{\cite{braunUsingThematicAnalysis2006}}. These were iteratively refined via multiple discussions with the full study team. Our final review of the codes ensured adequate coverage over all participants. The final set of codes and themes are shown below.

\begin{table}[ht!]
    \centering
    \begin{tabular}{p{0.9\textwidth}}
        \hline
        \textbf{Theme} / Code \\
        \hline
         \textbf{Participants' baseline understanding of (non-privacy)
         sources of noise shaped their perceptions of privacy noise} (\S~\ref{subsec:findings-general-noise})  \\
         $\cdot$ Participants characterized global pageview counts as not exact but reasonable estimates  \\
         $\cdot$ Many factors impact reliability of pageview counts (e.g. uniqueness, redirects, bots, translations) \\
         $\cdot$ Many said they would consider data provenance, comprehensiveness, and processing \\
         \hline
         \textbf{Participants more easily grasped the rounding process, but could develop analyses more effectively using the DP-protected data} (\S~\ref{subsec:findings-comfort-DP-rounding}) \\
         $\cdot$ Rounding seems to have larger impact on accuracy and provide less information than DP  \\
         $\cdot$ Rounding feels easier to conceptualize and communicate \\
         $\cdot$ DP feels easier to analyze because of its statistical properties, such as unbiasedness and Gaussian distribution \\
         $\cdot$ DP would be more useful than rounding in terms of aggregation, confidence, and thresholding \\
         \hline
         \textbf{Participants readily used simple uncertainty metrics from the documentation but struggled to compute confidence intervals across multiple perturbed data points} (\S~\ref{subsec:findings-simple-metrics-documentation}) \\
         $\cdot$ Participants relied on the provided documentation, especially visuals, accuracy metrics, and guidance on use, when completing tasks \\
         $\cdot$ Most participants provided the correct 100\% confidence interval for Rounding task 1 and 95\% confidence interval for DP Task 1.  \\
         $\cdot$ Several felt hesitant or confused about how to approach Rounding Task 2 and Rounding Task 3 due to the limits of rounding \\
         $\cdot$ Many participants incorrectly believed that when averaging noise over multiple data points for DP Task 2, the error would stay the same or increase in scale. \\
         $\cdot$ Many participants outlined a plausible simulation procedure to estimate the answer for DP Task 3. \\
         \hline
         \textbf{Participants had mixed understanding of both methods' effectiveness at protecting privacy} (\S~\ref{subsec:findings-privacy-perceptions}) \\
         $\cdot$ Many felt it was hard to compare the privacy protections of both methods. \\
         $\cdot$ Several were unsure about the privacy implications of publishing pageviews data and questioned the need for privacy protections.
         $\cdot$ Some said DP seemed to be more effective because of its statistical properties of random noise and unbiased estimates \\
         $\cdot$ Several incorrectly connected their perceived higher utility of DP to weaker privacy protections, and the lower accuracy of rounding to stronger privacy protections \\
         \hline
         \textbf{Participants had contextually-dependent preferences for which dataset to use when communicating their results} (\S~\ref{subsec:findings-communicating}) \\
         $\cdot$ For non-technical audiences or short presentations, participants felt that rounding was easier to explain. \\
         $\cdot$ For technical audiences, participants preferred to communicate using DP because of its rigorous statistical properties \\
         $\cdot$ If using DP, participants would spend additional time to understand the process themselves and develop communication techniques for their specific audiences \\
         $\cdot$ Participants suggested several communication techniques, including visuals, scenarios, comparisons, examples, and using rounding as an anchor \\
         \hline
    \end{tabular}
    \caption{Codebook that resulted from our thematic analysis of task-based interviews with data users.}
    \label{tab:codebook}
\end{table}

\end{document}